\documentclass[a4paper,fleqn,usenatbib]{mnras}

\usepackage[T1]{fontenc}
\usepackage{ae,aecompl}

\usepackage{graphicx}	
\usepackage{amsmath}	
\usepackage{amssymb}	

\title[Constant CN/HCN ratio in galaxies]{A nearly constant CN/HCN line ratio in nearby galaxies: CN as a new tracer of dense gas}

\author[C. D. Wilson et al.]{
Christine D. Wilson$^{1}$\thanks{E-mail: wilsoncd@mcmaster.ca},
Ashley Bemis$^{1,2}$,
Blake Ledger$^1$,
and Osvald Klimi$^{1}$ 
\\
$^{1}$Department of Physics and Astronomy, McMaster University, 1280
Main St. W., Hamilton, Ontario L8S 4M1, Canada\\
$^{2}$Leiden Observatory, Leiden University, PO Box 9513, 2300 RA Leiden, The Netherlands\\
}

\date{Accepted XXX. Received YYY; in original form ZZZ}

\pubyear{2022}

\begin{document}
\label{firstpage}
\pagerange{\pageref{firstpage}--\pageref{lastpage}}
\maketitle

\begin{abstract}
We investigate the relationship between CN $N=1-0$ and HCN $J=1-0$ emission on scales from 30 pc to 400 pc using ALMA archival data, for which CN is often observed simultaneously with the CO $J=1-0$ line. In a sample of 9 nearby galaxies ranging from ultra-luminous infrared galaxies to normal spiral galaxies, we measure a remarkably constant CN/HCN line intensity ratio of $0.86 \pm 0.07$ (standard deviation of 0.20). This relatively constant CN/HCN line ratio is rather unexpected, as models of photon dominated regions have suggested that HCN emission traces shielded regions with high column densities while CN should trace dense gas exposed to high ultraviolet radiation fields. We find that the CN/HCN line ratio shows no significant correlation with molecular gas surface density, but shows a mild trend (increase of $\sim 1.3$ per dex) with both star formation rate surface density and star formation efficiency (the inverse of the molecular gas depletion time). Some starburst and active galactic nuclei show small enhancements in their CN/HCN ratio, while other nuclei show no significant difference from their surrounding disks. The nearly constant CN/HCN line ratio implies that CN, like HCN, can be used as a tracer of dense gas mass and dense gas fraction in nearby galaxies. 
\end{abstract}

\begin{keywords}
galaxies: ISM -- galaxies: starburst -- ISM: molecules -- ISM:
abundances -- galaxies: abundances
\end{keywords}



\section{Introduction}
\label{sec:intro}

The relationship between the interstellar medium (ISM) and the rate at which stars form is critical for understanding galaxy formation and evolution. On global scales, galaxies at a wide range of redshifts follow the Kennicutt-Schmidt relation, which is often expressed as a correlation between
the global gas mass and the global rate of star formation \cite[e.g.,][]{kennicutt1998,tacconi2013,delosreyes2019,kennicutt2021}.
Galaxies in the nearby universe are also seen to obey a resolved Kennicutt-Schmidt relation, where the surface density of star formation correlates with the gas
surface density \citep[e.g.,][]{bigiel2008,schruba2011,lin2019ApJ...884L..33L}.

Over the past two decades, particular attention has been paid to the nearly linear correlation between dense molecular gas (typically traced by emission from HCN) and star formation rate \citep[e.g.,][]{gao2004,usero2015,gallagher2018}. The observed correlation with dense molecular gas even extends to the scale of individual star-forming cores inside giant molecular clouds in our own Milky Way \citep[e.g.,][]{wu2005,shimajiri2017}. Since regions of young star and star
cluster formation in Galactic molecular clouds are observed to be strongly spatially correlated with the densest clumps and cores inside those clouds \citep[e.g.,][]{lada1991,shimajiri2017}, such a correlation between
dense gas mass and star formation rate is to be expected. 
On extragalactic scales, resolved observations of nearby galaxies \citep[e.g.,][]{usero2015,gallagher2018,bemis2019}
demonstrate a strong correlation of star formation rate with dense gas surface density, although with indications of changing physical conditions in the central $\sim 1$ kpc regions. 
In several cases, dense gas tracers 
are found to be much more abundant in the irradiated ISM in the centres and nuclei of starburst galaxies \citep{meier2015,harada2018new,saito2022arXiv220706448S}.  
The dependence of the HCN emission on the gas temperature measured in the Integral Shaped Filament in Orion by \citet{hacar2020} further illustrates the complexity in directly relating HCN luminosity to dense gas mass.

In recent years, CN has been increasingly studied in nearby galaxies, beginning with spectral line surveys and observations of some of the brightest starbursts and active galactic nuclei \citep{walter2017, wilson2018, nakajima2018,harada2018new,takano2019PASJ...71S..20T,cicone2020A&A...633A.163C,cruz-gonzalez2020MNRAS.499.2042C,martin2021A&A...656A..46M,saito2022arXiv220706448S}. 
It is detected in a wide variety of environments, including local mergers and merger remnants \citep{ueda2017PASJ...69....6U,ueda2021ApJS..257...57U,ledger2021} and centrally concentrated in early type galaxies \citep{young2021ApJ...909...98Y,young2022ApJ...933...90Y}. It has been used to measure the $^{12}$C/$^{13}$C isotope ratio in 3 nearby starbursts \citep{tang2019A&A...629A...6T} and to study chemical evolution around the M83 circumnuclear ring \citep{harada2019m83}.
%
Enhanced CN emission has been observed in galactic outflows as well, where molecular dissociation and UV radiation impact the gas chemistry \citep{cicone2020A&A...633A.163C,saito2022arXiv220706448S}. 

Recent maps of individual Galactic molecular clouds  \citep{watanabe2017,gratier2017,kauffmann2017,bron2018,barnes2020} have shown that emission from dense-gas tracers such as HCN and CN can be detected over wide areas. 
 For example, \citet{kauffmann2017} 
find that half of the HCN emission in the Orion A molecular cloud comes
from regions with $A_v > 6$~mag and characteristic densities above $\sim 10^3$ cm$^{-3}$, significantly smaller than its critical density of $10^6$ cm$^{-3}$. 
Although \citet{kauffmann2017} do not derive a characteristic
density for CN, their data suggest it is likely to have a similar value to HCN.
Studies using radiative transfer models of clouds with a range of densities have also demonstrated that significant
HCN emission can be produced at densities $> 10^4$ cm$^{-3}$ \citep{shirley2015,leroy2017}, rather than exclusively at densities above $10^6$ cm$^{-3}$. 


\citet{bron2018} use a clustering analysis of maps of 5
molecular lines for the Orion B 
molecular cloud to show that CN emission is linked with UV-illuminated gas. 
They find that the CN abundance increases by an order
of magnitude over a similar range in  the UV-field to density ratio, $G_{\rm o}/n_{\rm H}$,
and then stays roughly constant for $G_{\rm o}/n_{\rm H} > 10^{-2}$. Using the same dataset and a Principal Component Analysis of 12 lines, \citet{gratier2017}
find that both CN and HCN show a
significant positive contribution to the principal component associated with the strength of the UV radiation field. 
In the OMC1-OMC3 region of Orion, \citet{brinkmann2020} 
find that CN correlates strongly with other high density tracers such as HCN, as well as with regions with strong UV illumination. Their work suggests that the CN/HCN line ratio may not be a reliable tracer of UV-illuminated gas.


\citet{wolfire2022} present a detailed overview of photon dominated regions (PDRs) 
including physical mechanisms, models, and comparison to observations. 
Constant density, plane-parallel PDR models 
predict that HCN is found at relatively high
column densities in the interior of clouds, where it is shielded from
UV radiation \citep[e.g.,][]{meijerink2005,meijerink2007}. In contrast, CN is expected to have a 
higher abundance in the outer layers of clouds \citep[e.g.,][]{boger2005}. From these models, we would
expect CN to be a tracer of UV-irradiated dense gas, while HCN traces
dense gas in shielded regions. However, there are many difficulties in making precise comparisons of models and observations \citep{wolfire2022}, of which the mixing of different components in the beam and the complicated geometry of the gas and the radiation fields, are particularly relevant for extragalactic observations. 

On the scale of entire molecular clouds, \citet{levrier2012A&A...544A..22L} 
presented one of the first attempts to combine simulations of a turbulent, structured ISM with a PDR code. They used a simulation without star formation and applied the Meudon PDR code on all gas with densities of 20-9000 cm$^{-3}$ exposed to the general interstellar radiation field. 
However, their CN abundances were a factor of 10 less abundant than observed, perhaps providing an early indication of the difficulties in modelling nitrogen chemistry. 

The most detailed recent work on modelling CN and HCN has been carried out in protoplanetary disks \citep{chapillon2012,arulanantham2020,bergner2021ApJS..257...11B}. Although the physical scales and structure are very different from what is observed in galaxies, this work provides insights into the mechanisms shaping the HCN and CN emission in PDRs and the dominant formation pathways for CN and HCN in protoplanetary disks. When H$_2$ is most abundant, CN formation is driven by the availability of H$_2^*$ molecules that are vibrationally excited by far-ultraviolet (FUV) photons \citep{arulanantham2020,bergner2021ApJS..257...11B}. HCN can then form from reactions of CN with H$_2$. The back reaction (destruction of HCN by UV photons) plays an important role in setting the total CN abundance in the disk. \citet{arulanantham2020} find that submillimeter CN fluxes are anti-correlated with the FUV continuum flux; in other words, although CN may be formed more efficiently when the FUV field is strong, it is destroyed more efficiently as well.

If a higher CN abundance is produced by strong UV radiation fields, we might expect the relative HCN abundance to be reduced where CN is abundant. 
With a combined study of CN, HCN, and CO (as well as the star formation rate) in a large sample of nearby galaxies spanning a wide range of star formation rate surface densities, we can measure the total amount of molecular gas (from CO), the amount of cold, dense molecular gas (from HCN), and the amount of UV-irradiated dense gas (from CN). In particular, we might expect that the CN/HCN line ratio would correlate with the star formation rate surface density, which we can take as a probe of the strength of the UV radiation field.

We started this project with the goal of understanding how
UV radiation from recent star formation affects the properties of dense gas. Unexpectedly, we found that the CN/HCN line ratio does not appear to be strongly affected by the star formation rate surface density and instead is extremely constant over nearly 3 orders of magnitude in $\Sigma_{\rm SFR}$ (Fig.~\ref{fig:ratio_vs_intensity}). \S~\ref{sec:obs} discusses the sample selection, data calibration and imaging. \S~\ref{sec:corr} presents the CN/HCN line ratio and discusses the implications for PDR models, as well as the promise of CN as a dense gas tracer. \S~\ref{sec:concl} presents our conclusions and a brief discussion of the potential for the CN $N=1-0$ line as a new tracer for dense gas in galaxies, one that can be observed simultaneously with the CO $J=1-0$ line using the Atacama Large Millimeter/submillimeter Array (ALMA).

\section{Sample and data reduction}
\label{sec:obs}

\subsection{Archival sample selection}
\label{subsec:sample}

\begin{table*}
	\centering
	\caption{Global properties of the 9 galaxies in the CN-HCN sample}
	\label{table:physical}
	\begin{tabular}{lcccccccl} 
		\hline
		Galaxy &
                $\log L_{\rm IR}$\footnotemark[1] 
                & $\log SFR$\footnotemark[2] 
                & $\log M_{\rm mol}$\footnotemark[3] 
                & $\log L_{\rm 2-10keV}$\footnotemark[4] 
                & $\log M_*$\footnotemark[2] 
                & Diameter\footnotemark[5] 
                & Distance\footnotemark[6]
                & Classification \\
                & (L$_\odot$) & (M$_\odot$ yr$^{-1}$) & (M$_\odot$) & (erg s$^{-1}$) & (M$_\odot$) & (kpc) & (Mpc) \\
		\hline
IRAS 13120-5453  & 12.29 &  2.46 & 9.8  
&  42.2  &  11.5  &  36  & 134 & ULIRG, AGN
\\
NGC 3256  & 11.75 &  1.92 & 9.9 
&  $< 41.8$  &  11.1 & 49  & 44 & LIRG, AGN
\\
VV 114  & 11.68 &  1.85 & 9.8 
&  $<42.5$  &  11.2 &  35  & 81 & LIRG, AGN
\\
NGC 7469  & 11.60 & 1.82 & 9.8
&  43.3  &  10.97  & 29  & 66.4 & LIRG, Seyfert 1.2
\\
NGC 1808  & 10.29 &  0.47  & 9.0 
&  39.6  &  9.86  & 15 
& 7.8 & Seyfert 2, starburst 
\\
M83  & 10.33 &  0.58 & 9.7
&  ... & 10.37   & 18 
& 4.7  &  starburst
\\
Circinus  & 10.08 &  0.5-0.9 & 9.2 
& 42.8 &  10.98  & 8 
& 4.2  &  Seyfert 2 
\\
NGC 3627  & 10.32 & 0.40  
& 9.6 & ... & 10.57 &  25 & 9.4 & LINER Seyfert 2 
\\
NGC 3351  & 9.77 &  0.01
& 8.9 &  ...  & 10.22  &  20 & 9.3 & starburst \\
		\hline
	\end{tabular}
	\\
\footnotemark[1] Infrared luminosities from \citet{sanders2003}
adjusted to luminosity distance given here, except for Circinus, where
we estimate $L_{IR}$ = 
$L_{FIR} + 0.15 $ with $L_{FIR}$ from \citet{shao2018}.
\\
{\footnotemark[2] For Circinus, star formation rate and stellar mass from \citet{for2012MNRAS.425.1934F}. For IRAS 13120, NGC 3256, and VV114, star formation rate calculated from $L_{IR}$ using the relation from \citet{kennicutt2012} and stellar masses from 2MASS K-band luminosity \citep[e.g., ][]{howell2010ApJ...715..572H}. Stellar masses and star formation rates from  \citet{leroy2019ApJS..244...24L} for the other 5 galaxies; note that NGC 7469, NGC 1808, and M83 may suffer from saturation and contamination by stars. \\
\footnotemark[3] Molecular gas masses using
 the ULIRG CO-to-H$_2$ conversion factor for the first 5 galaxies and the four-times-larger Milky Way conversion factor for the last 4 galaxies. Original CO data for galaxies (in order listed) from: \citet{sliwa2017ApJ...840L..11S}, \citet{sakamoto2014}, \citet{yamashita2017ApJ...844...96Y} , \citet{papadopoulos2012MNRAS.426.2601P}, \citet{salak2014PASJ...66...96S}, \citet{crosthwaite2002AJ....123.1892C}, \citet{curran2008MNRAS.389...63C}, \citet{leroy2021ApJS..257...43L}, \citet{leroy2009AJ....137.4670L}. 
\\
\footnotemark[4] AGN X-ray luminosity for NGC 1808 from
\citet{combes2019A&A...623A..79C} and for Circinus from \citet{uematsu2021ApJ...913...17U}; all other galaxies from \citet{yamada2021ApJS..257...61Y}, with  upper limits  assuming $N_H = 10^{25}$ cm$^{-2}$.\\
\footnotemark[5] RC3 diameters from the NASA/IPAC Extragalactic Database (NED); for IRAS 13120-5453 the diameter is from 2MASS and for VV 114 the diameter is from ESO-Uppsala. \\}
\footnotemark[6] NGC 7469: \citet{ganeshalingam2013} 
with $H_o = 70.7$ km s$^{-1}$ Mpc$^{-1}$. 
NGC 1808: \citet{sorce2014}. 
Circinus: \citet{tully2009}. 
NGC 3627, NGC 3351: \citet{freedman2001}. 
M83: averaged from \citet{radburn2011} and \citet{saha2006}. 
Remaining distances are WMAP 5-year cosmology ($H_o =
70.5$ km s$^{-1}$ Mpc$^{-1}$, $\Omega = 1$, $\Omega_{\rm m} = 0.27$) in CMB frame. All quantities in this table are adjusted to these distances.\\
\end{table*}

\begin{table*}
	\centering
	\caption{ALMA projects with CO, CN, HCN, and 93 GHz continuum observations of
          nearby galaxies}
	\label{table:sample}
	\begin{tabular}{lccc} 
		\hline
Galaxy & CO and CN & HCN and continuum & phase centre \\
 & project code &  project code & coordinates (J2000) \\
		\hline
IRAS 13120-5453 & 2015.1.00287.S & 2013.1.00379.S & 13:15:06.3419 55:09:22.8770 \\
NGC 3256 & 2011.0.00525.S  & 2015.1.00993.S & 10:27:51.2300 -43:54:16.6000 \\ 
VV 114 & 2011.0.00467.S  & 2013.1.01057.S & 01:07:47.2080 -17:30:24.8400 \\ 
NGC 7469 & 2013.1.00218.S & 2012.1.00165.S & 23:03:15.6400 +08:52:25.8000 \\
NGC 1808 & 2012.1.01004.S & 2013.1.00911.S & 05:07:42.3430 -37:30:46.9800 \\
M83 & 2011.0.00772.S  & 2015.1.01177.S & 13:37:00.9190 -29:51:56.7400 \\ 
Circinus & 2013.1.00247.S & 2015.1.01286.S & 14:13:09.9060 -65:20:20.4698 \\
NGC 3627 & 2015.1.01538.S & 2013.1.00634.S & 11:20:15.3903 +12:59:44.4766 \\
NGC 3351 & 2013.1.00885.S & 2013.1.00634.S & 0:43:57.7330 +11:42:12.9996 \\
		\hline
	\end{tabular}
\\
Note: The galaxy sample was identified in a search of the ALMA archive on 2017 August
7; see \S\ref{subsec:sample} for more details. 
\end{table*}

For our study, we needed to identify  a
sample of galaxies observed with ALMA in  both
CO $J=1-0$ and the traditional dense gas tracer, HCN $J=1-0$.  Since the CN $N=1-0$ lines lie $\sim 2$~GHz below the CO line, most ALMA
extragalactic projects that target CO $J=1-0$ also serendipitously
observe the CN lines simultaneously. HCN is at a
significantly lower frequency and so must be observed separately; these data sets also provided us with 93 GHz continuum fluxes that we can use to measure the star formation rate (see \S\ref{subsec:physical}).

We searched the ALMA archive on 2017 August 7 for publicly available
data on galaxies with redshifts $<0.05$. At this redshift,  1$^{\prime\prime}$ corresponds to 1 kpc, which provides a good tradeoff between resolution and sensitivity for faint emission lines. For HCN, we selected
all data with observed frequencies less than 88.7 GHz, while for CN
and CO, we selected all projects observed in Band 3 with
frequencies greater than 107.5~GHz. These searches produced two long
lists of observations containing everything from protostars to high
redshift galaxies. From this list, we identified a total of 12 galaxies with resolved detections of CO, CN, and HCN.
We opted to
exclude Arp~220 from our analysis because of the extremely high resolution of the data \citep{sakamoto2017,scoville2017}. We similarly excluded NGC~253 from our
analysis because its bright extended emission would have required us
to combine data from both the 12m and compact arrays \citep[e.g.,][]{leroy2015}.
Finally, although VV219 showed some extended emission for HCN and CN,
it was not detected in 93 GHz continuum and thus it was
excluded from further analysis. 

Our final sample of 9 galaxies
(Table~\ref{table:physical}) is quite heterogeneous, and
includes galaxies with a prominent active galactic nucleus (AGN) or
Seyfert nucleus
(Circinus, NGC1808, NGC3627, NGC 7469), luminous and ultraluminous
infrared galaxies 
(U/LIRGs: IRAS 13120-5453, NGC3256, VV114), galaxies with a starburst
nucleus (M83, NGC 1808), and spiral galaxies (NGC3627,
NGC3351). 

\subsection{Calibration and spectral imaging}
\label{subsec:calibration}

\begin{table*}
	\centering
	\caption{Galaxies and their image processing parameters}
	\label{table:cubes}
	\begin{tabular}{lllllllllllr} 
		\hline
& & & & \multicolumn{2}{c}{uvtaper applied} & \multicolumn{4}{c}{Sensitivity per channel\footnotemark[2]} & \multicolumn{2}{c}{Binned} \\
		Galaxy & beam & $\Delta V$\footnotemark[1]  & uvrange cut
                & (CO, CN) & (HCN, 93GHz)  & CO
                & CN  &    HCN  & 93
                GHz  & \multicolumn{2}{c}{pixels\footnotemark[3]} \\
 & (arcsec) & (km s$^{-1}$) &
             (k$\lambda$) & \multicolumn{2}{c}{
                (arcsec $\times$ arcsec, PA in
                  degrees)} & \multicolumn{4}{c}{ (mJy beam$^{-1}$)}  &  (arcsec) &  (pc) \\ 
		\hline
IRAS 13120 & 1.1 & 20 & $> 15$  &
0.8$\times$0.9, -23 & 0.0$\times$0.8, 71 & 1.4 & 1.2 & 0.95 & 0.080 & 0.6&  390 \\ 
NGC 3256 & 2.2 & 26.43 & $> 5$ &
0.0$\times$1.2, -69
& 0.0$\times$1.2, 88 & 0.85 & 0.6 & 0.27 & 0.050 & 0.9 & 192 \\ 
VV 114 & 2.3 & 20 & $>$ 9.5 & 0.0$\times$1.4, -86 & 1.5$\times$1.7, 83 & 1.75 & 1.4 & 0.67 & 0.055 & 0.9 & 353 \\ 
NGC 7469 & 0.95 & 20 & $>18$ &
0.0$\times$0.6, -47 & ...  & 0.5 & 0.5 & 0.27 & 0.015 & 0.6 & 193 \\ 
NGC 1808 & 3.75 & 41.28 & ... & 2.4$\times$3.0, -82 & 0.0$\times$2.0, -82 & 3.1 & 1.7 & 0.70 & 0.13 & 1.5 & 57 \\ 
M83 & 2.1 & 10 & $>$ 9  & 0.0$\times$1.7, -84 & 0.0$\times$1.4, 80 & 6.9 & 4.0 & 0.80 & 0.055  & 1.2 & 27 \\ 
Circinus & 3.0 & 20 & ...  & 0.0$\times$1.5, 72 & ... & 2.4 & 1.7 & 0.28 & 0.067  & 1.5 & 31 \\ 
NGC 3627 & 3.6 & 20 & ... & 1.7$\times$2.6, 54 & 0.0$\times$2.0, 83 & 1.3 & 0.85 & 0.50 & 0.033 & 1.5 & 68 \\ 
NCG 3351 & 3.45 & 41.28 & $>$ 10 & 2.7$\times$2.8, 35 & 0.0$\times$1.9, 89 & 3.3 &0.8  & 0.55 & 0.015 & 1.8 & 81 \\ 
		\hline
	\end{tabular}
\\
\footnotemark[1] We typically used a value of 20 km/s except when CN or HCN
data were taken with lower spectral resolution, in which case this limited our velocity
resolution. See \S~\ref{subsec:calibration} for details. \\
\footnotemark[2] RMS noise levels used to set cleaning threshold; estimated from emission-free channels at start and end of dirty cube. \\
\footnotemark[3] Distances to the galaxies are given in Table~\ref{table:physical}.
\\
\end{table*}

The datasets used for this project are given in Table~\ref{table:sample}. For datasets taken in ALMA Cycle 0 (project codes 2011.0.*), we
downloaded the calibrated uvdata directly from the ALMA archive. 
For IRAS~13120-5450, calibrated uvdata were provided to us by the
project PI (K. Sliwa). For the remaining datasets
(Table~\ref{table:sample}), 
we downloaded the raw uvdata and calibration scripts from the ALMA
archive. We produced calibrated 
uvdata by running the provided calibration script  (``ScriptForPI.py'')
in the appropriate version of CASA \citep{mcmullin2007}.
A few galaxies had data from the
Atacama Compact Array available as well; we did not use these data in
our analysis. All further processing was done in various CASA 5 versions (5.1-5.4).


Continuum subtraction was done in the uv-plane using line-free
channels. For each galaxy, the baseline range was plotted to identify
the range of baseline lengths in common between the HCN and CO data sets.
Imaging was done using the CASA command \texttt{tclean} using the parameters
listed in Table~\ref{table:cubes}. We used Briggs weighting with
\texttt{robust}=0.5 and the deconvolve method set to \texttt{hogbom}, with mosaic gridding
used when needed. 
When only low spectral resolution data were
available for CN (NGC~1808, NGC~3351) or HCN (NGC~3256), that line was imaged in channel space with no
binning; the other two lines were then imaged using velocity widths 
matched to the output from the low spectral resolution cube. We adopted a mean rest frequency of 
113.490985~GHz for the brighter of the two groups of CN hyperfine lines; 
the fainter hyperfine group near 113.144190~GHz was also imaged \citep[see, e.g.,][]{ledger2021} but is not 
used in this analysis.

We first made a dirty cube of each line and measured the rms noise in
line-free channels. When necessary, we adjusted the \texttt{uvtaper} parameter
through trial and error to obtain a beam that was slightly
smaller than that of the  dataset with lower angular resolution. We
cleaned the images to a 2 sigma noise cutoff; occasionally we would
clean with a higher cutoff initially and follow with a second deeper
clean. All clean boxes were drawn interactively in each channel for
each galaxy.  The cleaned images were
smoothed to a common resolution using the CASA command \texttt{imsmooth} to
create angular and velocity resolution matched data cubes.

We made  93 GHz continuum images for each galaxy using all the line-free
channels from all the spectral windows of the HCN dataset. We first split
and rebinned the data in frequency to increase the imaging
speed and then used \texttt{tclean} to make a dirty and then a cleaned map
using multi-frequency synthesis. We used \texttt{imsmooth} to smooth the
cleaned continuum image to the same resolution as the line cubes.
Table~\ref{table:cubes} summarizes the imaging parameters for each
galaxy. 

\subsection{Further data processing}
\label{subsec:moreimaging}

\begin{figure*}
    \includegraphics[width=0.87\textwidth]{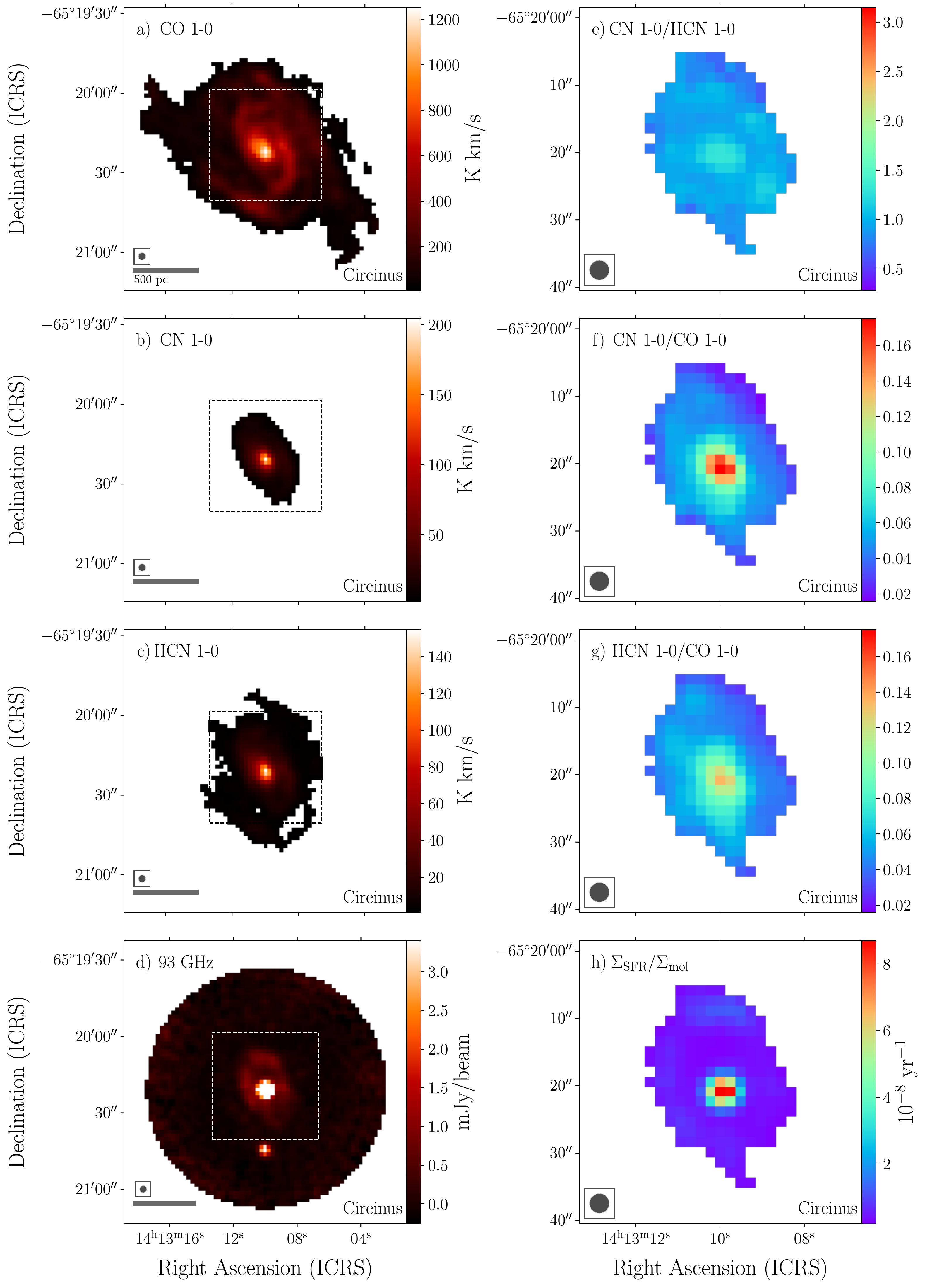}
    \caption{(Left column) Images of gas and star formation rate tracers in the Circinus galaxy: from top to bottom, CO(1-0), CN(1-0) (brighter hyperfine grouping), HCN(1-0), 93 GHz continuum. (Right column) Maps of line and continuum ratios in Circinus: from top to bottom, CN/HCN ratio, CN/CO ratio, HCN/CO ratio,  $\Sigma_{\rm SFR}/\Sigma_{\rm mol}$ ratio. The presence of a strong central AGN means that $\Sigma_{\rm SFR}$ and $\Sigma_{\rm SFR}/\Sigma_{\rm mol}$ are overestimated in the central pixels for this galaxy. Note that a smaller field of view is shown for right column (ratio images) compared to the left column (line and continuum). Only pixels with signal to noise greater than 4 in all lines and continuum are shown in the right column. The beam size is the same for all images and is indicated by the circle in the lower left corner of each image.
    }
    \label{fig:exampleimages}
\end{figure*}

For analyzing the line ratios, we binned the cleaned image cubes to have pixels
with roughly Nyquist sampling (Table~\ref{table:cubes}). Moment maps (and matching noise
maps) were made using the method described in \citet{sun2018}, with uncertainties calculated as described in Appendix~\ref{app:uncerts}. We added a
5\% calibration uncertainty in quadrature to the noise maps.
Finally, we mask
the images so that only pixels that are detected with a signal-to-noise of
at least 4 in all 3 lines, as well as in the radio continuum, are included for
further analysis. The resulting images and image ratios are shown in Fig.~\ref{fig:exampleimages} for the Circinus galaxy and in Figs.~\ref{fig:appendixfig10}-\ref{fig:appendixfig5} in Appendix~\ref{app:images} for the other 8 galaxies.

\subsection{Conversion to physical quantities}
\label{subsec:physical}

\begin{figure*}
	\includegraphics[width=\columnwidth]{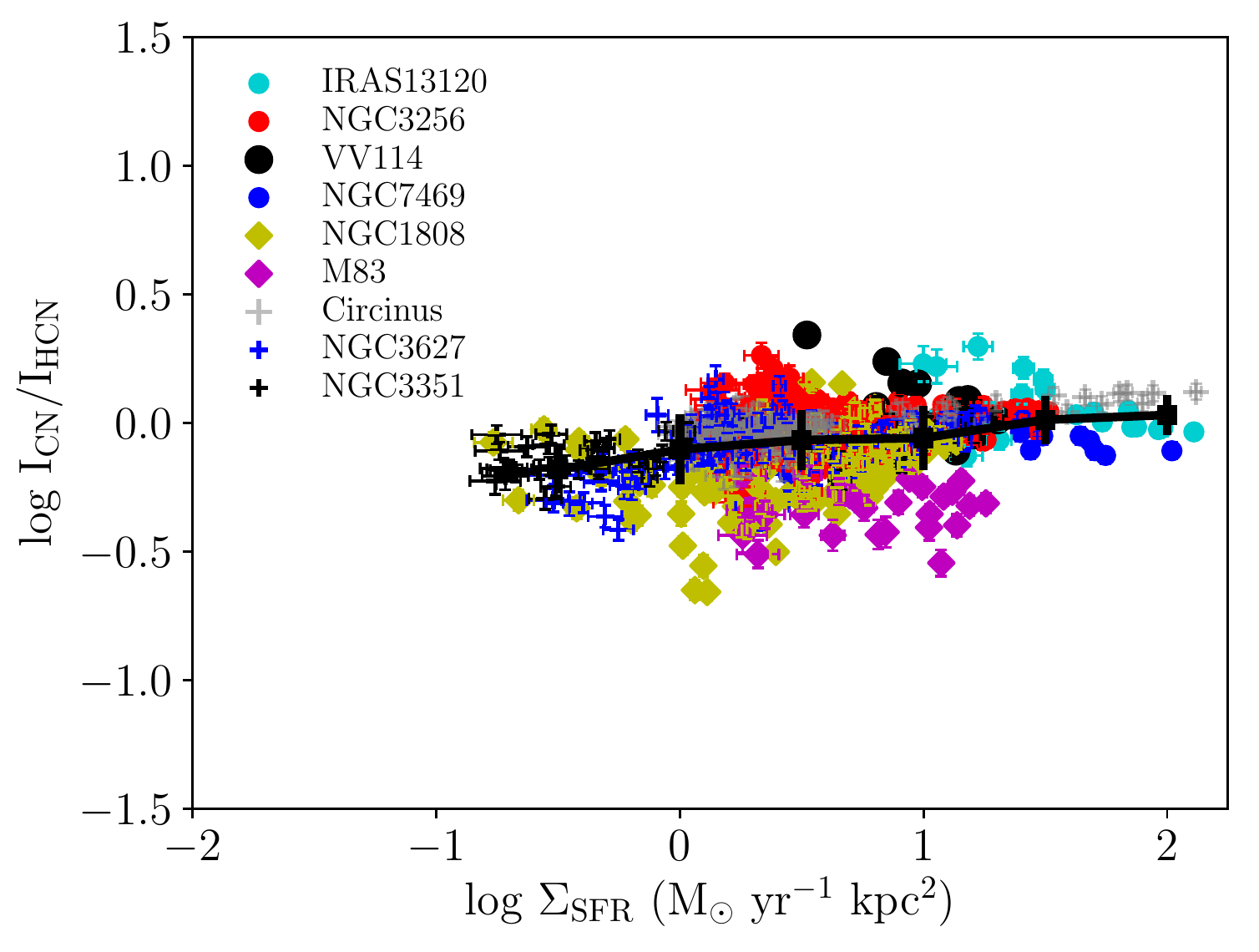}
	\includegraphics[width=\columnwidth]{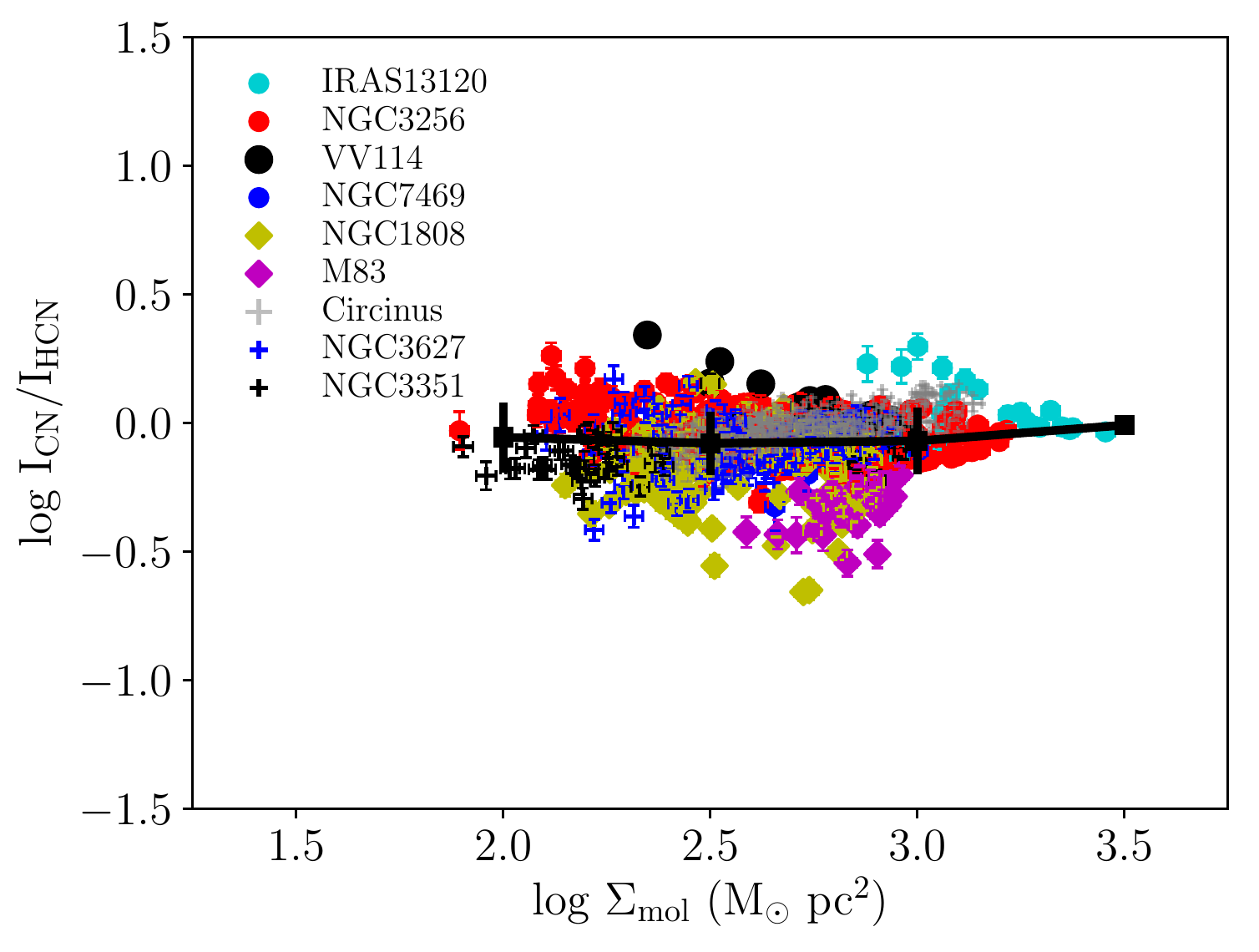}
    \caption{(left) The ratio of CN/HCN intensity plotted as a function of the
      star formation rate surface density for independent pixels
      in 9 nearby galaxies. The black symbols and line shows the mean and standard deviation in logarithmic bins spaced by 0.5. This ratio shows only a small increase over more than two orders of magnitude in star formation rate surface density. } The CN/HCN ratio is calculated from
      measurements of the intensities in units of K
      km s$^{-1}$. The star formation rate surface density is
      estimated from the 93 GHz continuum emission; no correction has
      been made for dust or synchrotron contamination (see \S~\ref{subsec:physical}). 
 Note that the brightest pixels in NGC~7469 and Circinus
      contain AGN nuclei and so the star formation rate surface
      density for these pixels will be an overestimate.
(right) The ratio of CN/HCN plotted as a function of the
      molecular gas surface density. The CN/HCN ratio  shows no 
      trends with gas surface density, even when the data are binned.
    \label{fig:ratio_vs_intensity}
\end{figure*}

To convert the CO integrated intensities to molecular gas surface
densities, we adopt a value for the CO-to-H$_2$ conversion factor of
$0.5\times 10^{20}$ H$_2$ cm$^{-1}$ (K km s$^{-1}$)$^{-1}$ 
and include a factor of 1.36 for helium.
This is the conversion factor that is typically used for U/LIRGs
\citep{downes1998} but is also a reasonable value for the inner
kiloparsec regions of our less extreme spiral galaxies \citep{sandstrom2013ApJ...777....5S}. The molecular
gas surface density, $\Sigma_{\rm mol}$ in M$_\odot$~pc$^{-2}$, is
then given by 
$\Sigma_{\rm
mol} = 1.36 \Sigma_{\rm H_2} = 1.088 I_{\rm CO(1-0)}$ where 
$I_{\rm CO(1-0)}$ is the  CO
integrated intensity in 
 K~km~s$^{-1}$. 
$\Sigma_{\rm mol}$ is the observed surface density; for a galaxy with
inclination $i$, the 
true surface density perpendicular to the plane of the galaxy will be smaller by a factor of $\cos i$. We do not correct for this inclination factor,
which is unknown for our most disturbed systems. In any case, the inclination
factor will cancel out for ratios of surface densities.
The observed molecular gas surface densities range from 300 to
 10$^4$ M$_\odot$ pc$^{-2}$ (see Fig.~\ref{fig:ratio_vs_intensity}). 

These high gas surface densities imply a high dust extinction and so we
require a star 
formation rate tracer that is insensitive to dust obscuration yet
still can be measured at arcsecond-scale resolution. 
Although methods such as H$\alpha$ imaging could be
used for the spiral galaxies in our sample, optical imaging cannot
probe the centers of the U/LIRGs in our sample. Thus, the most appropriate
star formation rate
tracer for our sample is the radio continuum
\citep{murphy2011}.
We therefore calculate the star formation
rate surface density using our 93 GHz continuum images, the pixel area in pc$^{2}$ (see Table~\ref{table:cubes}), and the
thermal-only formula from \citet{murphy2011},
 $$SFR  = 4.6 \times 10^{-28} T_e^{-0.45} \nu^{0.1} L_\nu$$ where $SFR$ is the star formation rate in $M_\odot$ yr$^{-1}$, $T_e$ is the electron temperature in units of $10^4$ K (we adopt $T_e = 10^4$ K), $\nu$ is the observed frequency in GHz, and $L_\nu$ is the thermal spectral luminosity in units of erg s$^{-1}$ Hz$^{-1}$.
These star formation rate surface densities range from 0.1-100 M$_\odot$ yr$^{-1}$ kpc$^{-2}$
(Fig.~\ref{fig:ratio_vs_intensity}). 
We make no correction for contamination of the 93 GHz emission from dust emission, which is estimated to contribute at most $\sim$10 percent of the emission for the 3 most luminous galaxies in our sample \citep[][]{wilson2019}. We also make no correction for contamination from synchrotron emission, which means that the star formation rate surface densities will be overestimated near AGN, particularly in NGC 7469 and Circinus but also possibly in other galaxies containing an AGN (Table~\ref{table:physical}). Finally, the true star formation rate surface density  perpendicular to the
galactic disk will be smaller by a factor of $\cos i$.

\section{Line ratios of the dense gas tracers CN and HCN}
\label{sec:corr}

\subsection{A constant CN/HCN line ratio}
\label{subsec:constLineRatio}

\begin{table*}
	\centering
	\caption{Average line ratios of CN, HCN, and CO in 9 galaxies}
	\label{table:ratios}
	\begin{tabular}{lccccccl} 
		\hline
		Galaxy & CN/HCN\footnotemark[1] & CN/HCN\footnotemark[1] & CN/CO\footnotemark[1]
                & HCN/CO\footnotemark[1] 
                & N 
                & beam FWHM & Classification \\
                & (center)\footnotemark[2] & (global)\footnotemark[4] & (global)\footnotemark[4] & (global)\footnotemark[4] & (pixels) & (pc) \\
		\hline
IRAS 13120-5453  & 0.98$\pm$0.03 &  1.19 $\pm$ 0.08  &  0.176 $\pm$ 0.015   &  0.164 $\pm$ 0.019
&  20& 715 & ULIRG, AGN
\\
NGC 3256  & 1.10$\pm$0.01\footnotemark[3] &  0.96 $\pm$ 0.02   &  0.045 $\pm$ 0.002 &  0.046 $\pm$ 0.001  
&  206 & 470 & LIRG, AGN
\\
VV 114  & 1.05$\pm$0.05 &  1.07 $\pm$ 0.08    &  0.035 $\pm$ 0.004  &  0.032 $\pm$ 0.002
&  21  & 900 & LIRG, AGN
\\
NGC 7469  & 0.81$\pm$0.03 &  0.87 $\pm$ 0.02   &  0.103 $\pm$ 0.006 &  0.118 $\pm$ 0.007
&  54 & 305 & LIRG, Seyfert 1.2
\\
NGC 1808  & 0.85$\pm$0.03  &  0.70 $\pm$ 0.02    &  0.061 $\pm$ 0.004  &  0.083 $\pm$ 0.003
&  112 & 142 & Seyfert 2, starburst 
\\
M83  & ... &  0.47 $\pm$ 0.02   &  0.046 $\pm$ 0.002 &  0.098 $\pm$ 0.003
&  25 & 47 &  starburst
\\
Circinus  & 1.28$\pm$0.02 &  0.94 $\pm$ 0.01   &  0.052 $\pm$ 0.002 &  0.054 $\pm$ 0.001
&  189  & 62 &  Seyfert 2 
\\
NGC 3627  & 0.76$\pm$0.02 &  0.82 $\pm$ 0.04   &  0.031 $\pm$ 0.001  &  0.040 $\pm$ 0.002
&  49  & 163 & LINER Seyfert 2 
\\
NGC 3351  & ... &  0.73 $\pm$ 0.02   &  0.064 $\pm$ 0.002 &  0.088 $\pm$ 0.002
&  34 & 155 & starburst \\
		\hline
	\end{tabular}
	\\
\footnotemark[1] Mean integrated intensity ratio and uncertainty on
the mean. \\
\footnotemark[2] Measured in the 5 pixels centered on the radio continuum peak pixel. \\
\footnotemark[3] For the northern starburst nucleus; the mean CN/HCN ratio for the southern AGN nucleus is 0.94$\pm$0.02.\\
\footnotemark[4] Across all detected pixels (see \S~\ref{subsec:moreimaging}). Note that the maps do not cover the full extent of the emission in the nearest galaxies (see Table~\ref{table:physical}).
\\
\end{table*}

\begin{figure*}
	\includegraphics[width=\columnwidth]{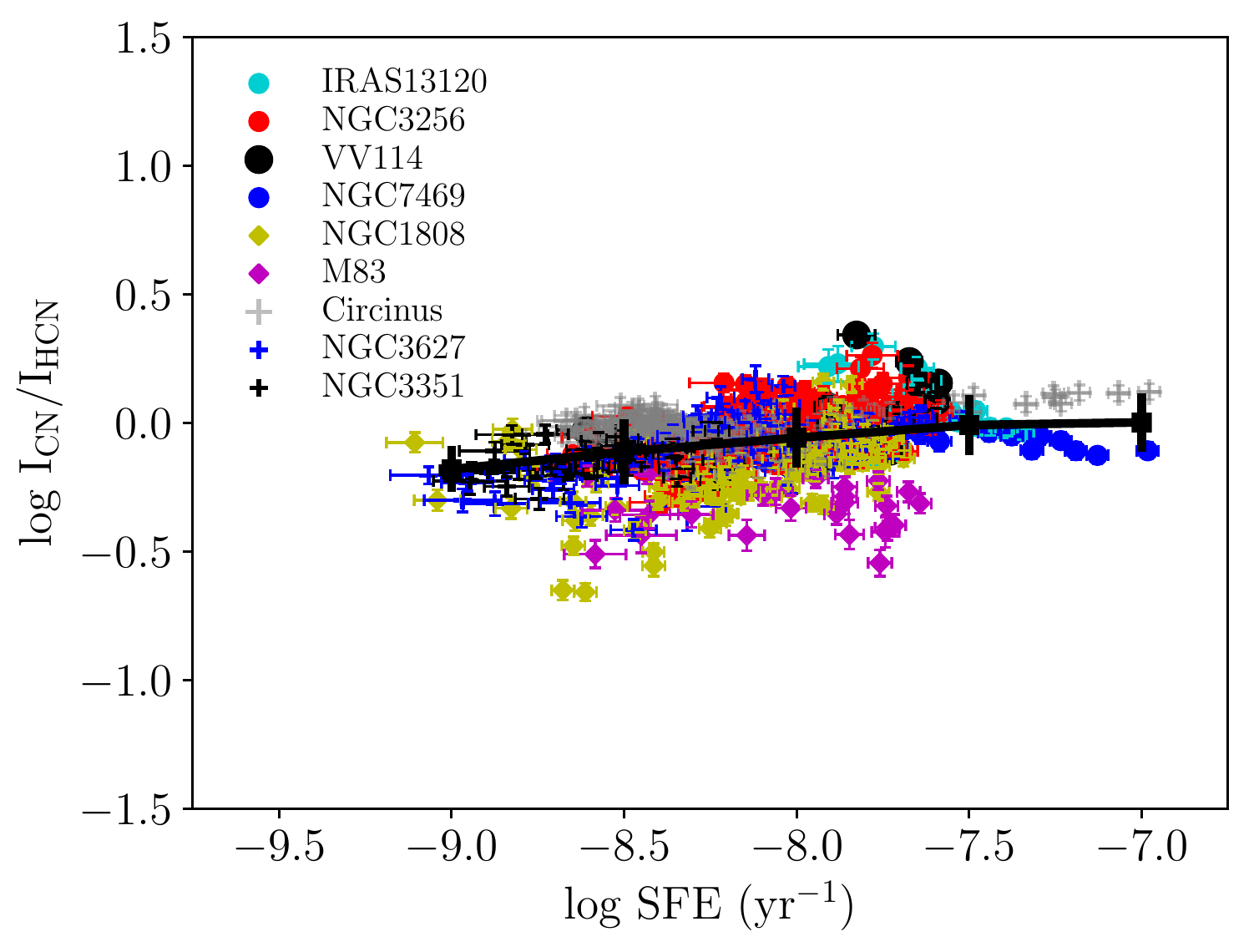}
	\includegraphics[width=\columnwidth]{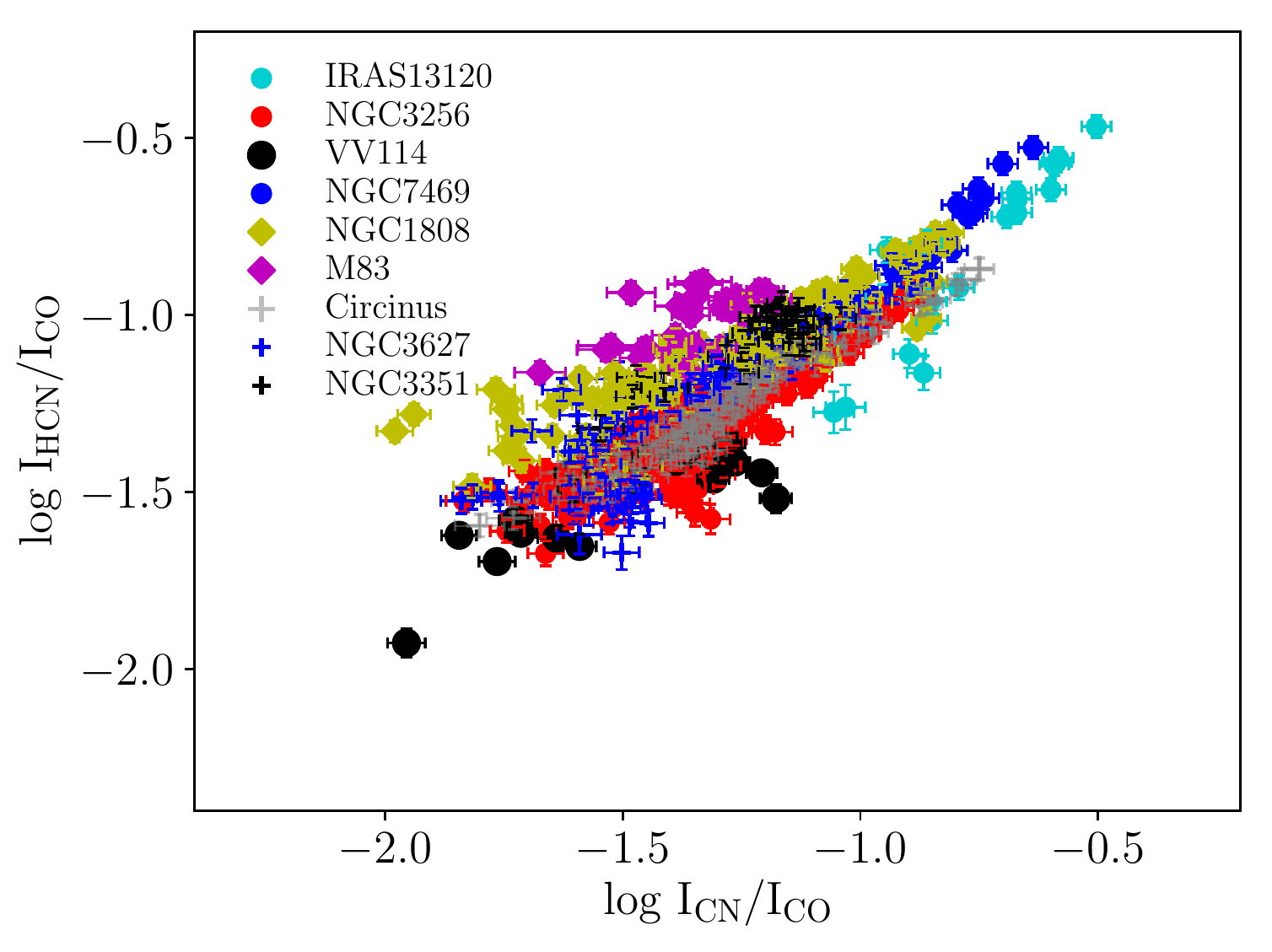}
    \caption{(left) The ratio of CN/HCN intensity is plotted as a function of the star formation efficiency (the ratio
      of the star formation rate surface density to the molecular gas
      surface density). The black symbols and line shows the mean and standard deviation in logarithmic bins spaced by 0.5. A mild increase in CN/HCN is seen with increasing SFE. Some mild trends are also 
      visible for individual galaxies, particularly 
    for NGC~1808. }The pixels with the largest $\Sigma_{\rm
      SFR}/\Sigma_{\rm mol}$ ratios are the
    AGN nuclei of NGC~7469 and Circinus. See Fig.~\ref{fig:ratio_vs_intensity} for
    more details. (right) The CN/CO ratio versus the HCN/CO ratio. The HCN/CO ratio is often used to measure the dense gas fraction \citep[e.g.,][]{jimenez2019}. The tight correlation of these two ratios implies that CN/CO can also be used to trace the dense gas fraction.
    \label{fig:ratio_vs_ratio}
\end{figure*}

\begin{figure*}
	\includegraphics[width=\columnwidth]{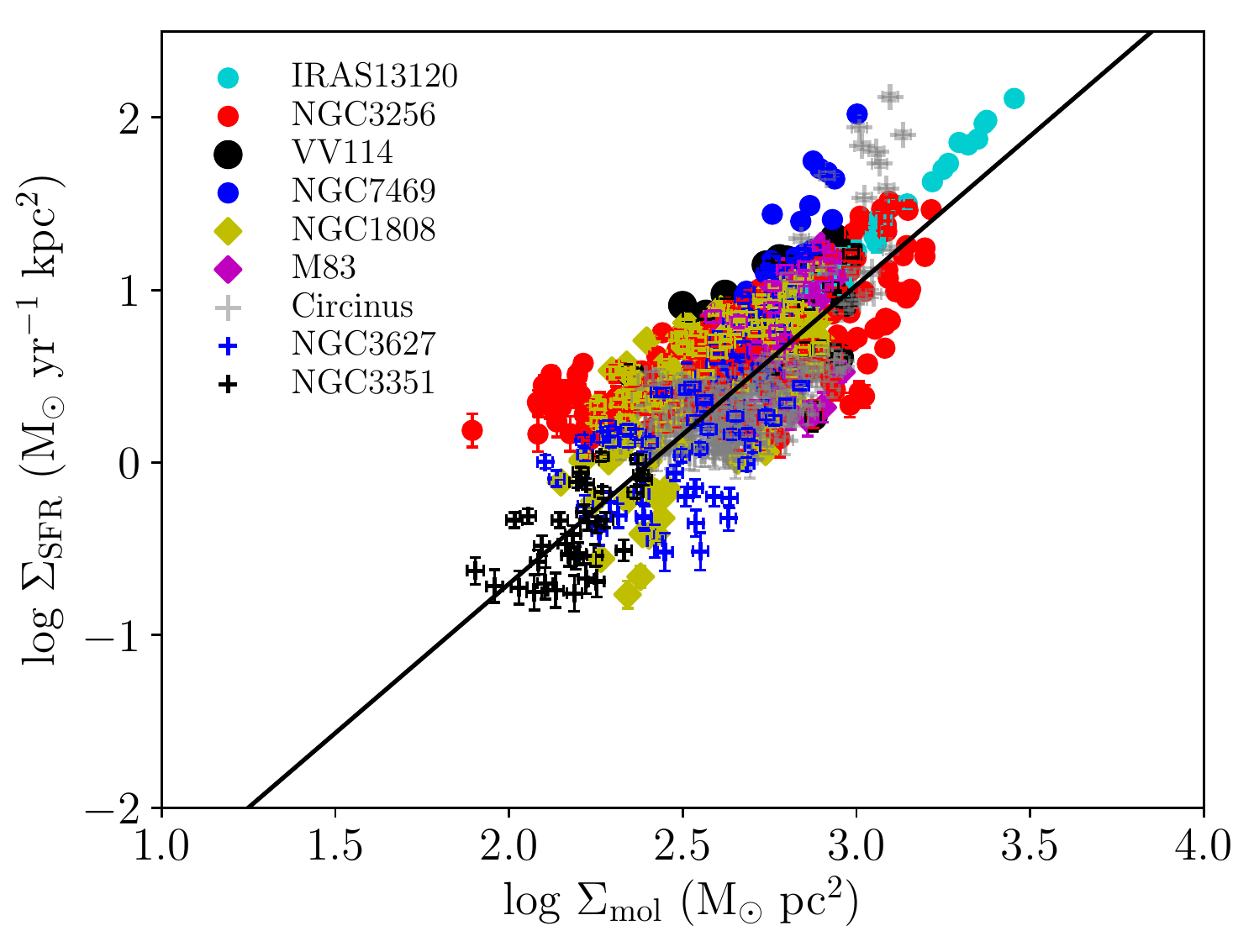}
	\includegraphics[width=\columnwidth]{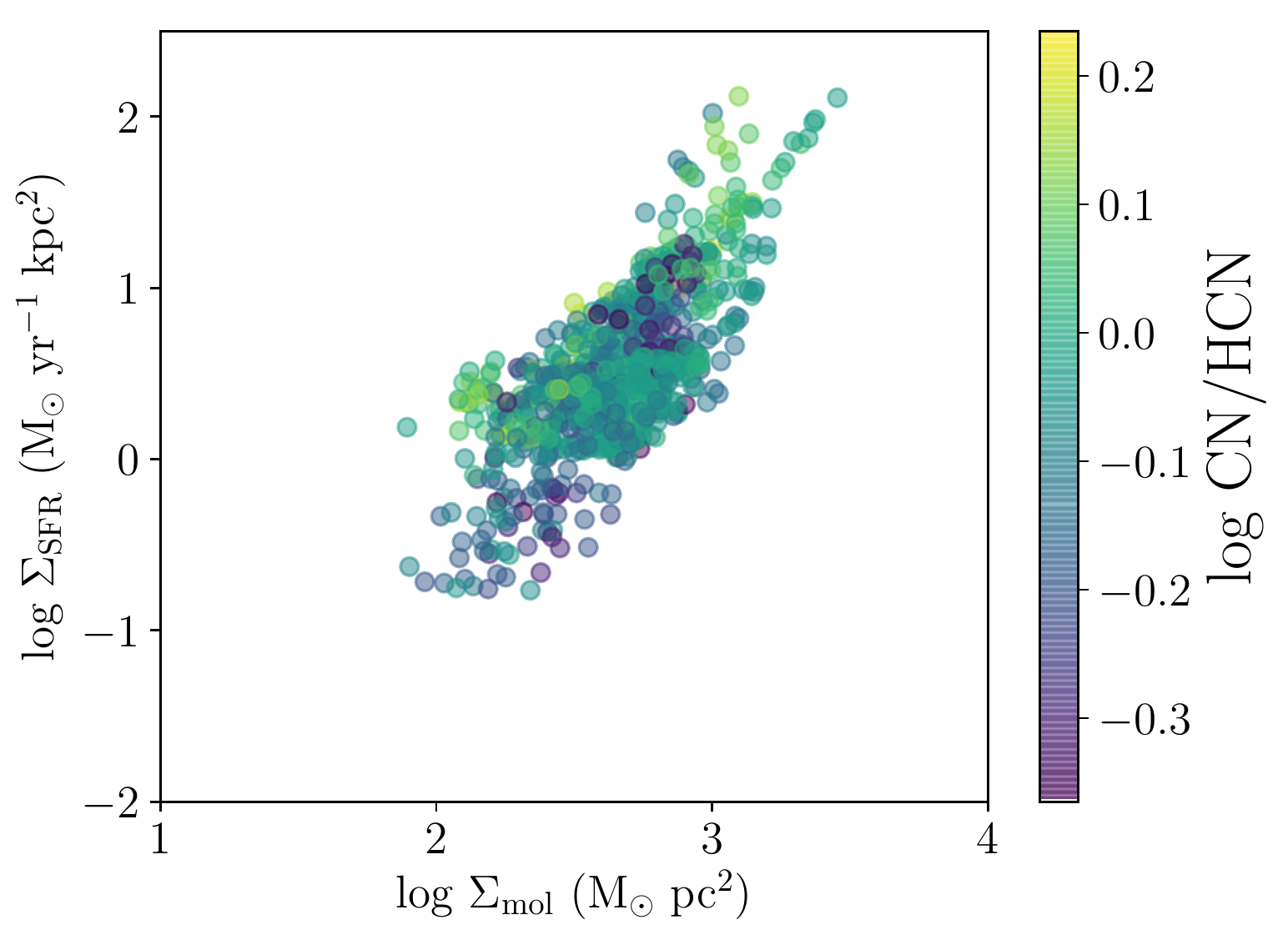}
    \caption{(left) The resolved Kennicutt-Schmidt relation between star formation rate and molecular gas surface densities for the 9 galaxies in our sample. The black line shows the fit to resolved data for U/LIRGS from \citet{wilson2019}. The data follow the trend quite well; the few apparently high $\Sigma_{\rm SFR}$ points lying above the relationship for NGC 7469 and Circinus correspond to central pixels containing emission from  bright AGN nuclei.  (right) The same Kennicutt-Schmidt relation with the points color-coded by the logarithmic value of the CN/HCN intensity ratio. There is no obvious trend of the CN/HCN ratio along or across the Kennicutt-Schmidt relation for these galaxies.}
    \label{fig:ksplots}
\end{figure*}

We calculate the CN/HCN line intensity ratio for each individual pixel after
converting the individual images to units of K km s$^{-1}$. 
Fig.~\ref{fig:ratio_vs_intensity} shows that the CN/HCN line 
ratio is remarkably constant across at least
2 orders of magnitude in both star formation rate surface density
and gas surface density. From the central
kiloparsec of nearby spirals to an ultraluminous infrared galaxy, and even near bright AGN, there
is remarkably little  variation in this line
ratio. 
Fig.~\ref{fig:ratio_vs_ratio} shows that the CN/HCN line ratio is also quite constant as a
function of the resolved star formation efficiency, SFE=$\Sigma_{\rm SFR}/\Sigma_{\rm mol}$.\footnote{This quantity has units of 1/time and is 
the inverse of the instantaneous depletion time of the molecular gas. It should not be confused with a true efficiency such as the efficiency per free fall time $\epsilon_{\rm ff}$ \citep[c.f.][]{wilson2019}.}
 In other words, increasing the ratio of UV photons to H$_2$
molecules (as traced by the SFE) by a factor of 30 does not produce a significant increase in
the CN/HCN line ratio. 

Giving equal weight per galaxy, the mean CN/HCN ratio in our sample is 
$ 0.86 \pm 0.07$ (with a standard deviation of 0.20). 
Comparing the individual mean ratios for each
galaxy shows a full range of only a factor of 2.5 (0.40 in log space), 
with M83 having the lowest average CN/HCN
line ratio and IRAS 13120-5453 the highest ratio  (Table~\ref{table:ratios}). 
If we bin the data shown in Fig.~\ref{fig:ratio_vs_intensity} and Fig.~\ref{fig:ratio_vs_ratio}, we see a mild trend of increasing CN/HCN ratio with $\Sigma_{\rm SFR}$ and $\Sigma_{\rm SFE}$, but no trend with $\Sigma_{\rm mol}$. Ignoring the highest surface density bin in each case (where contamination of the 93 GHz emission from the AGN in NGC 7469 and Circinus is present), the CN/HCN ratio increases by a factor of 1.6 for a factor of 100 change in $\Sigma_{\rm SFR}$ and by a factor of 1.5 for a factor of 30 change in SFE. 

We can compare this mean CN/HCN line ratio to previous results for Galactic molecular clouds.
In the \citet{kauffmann2017} study of the Orion A molecular cloud, the peak CN intensity is about 50\% that of HCN, somewhat smaller than the mean value of 0.86 obtained in our sample. Furthermore, the HCN emission in \citet{kauffmann2017} is more spatially extended than the CN emission, which implies that the CN/HCN luminosity ratio integrated over the entire cloud would be significantly smaller than the ratios see in Figure 2. \citet{barnes2020} also find the HCN emission to be more extended and brighter than the CN emission in the W49 molecular cloud.

Fig~\ref{fig:ksplots} shows the resolved Kennicutt-Schmidt relation between star formation rate surface density and molecular gas surface density. The data points scatter around the fit to five luminous and ultraluminous infrared galaxies (U/LIRGs) from \citet{wilson2019}. \footnote{IRAS 13120, NGC 3256, and NGC 7469 were part of that analysis, although the pixels were a factor of two larger than the ones used in this paper.} This starburst Kennicutt-Schmidt relation is offset from the one found in normal spiral galaxies \citep[e.g., ][]{kennicutt2021}. Even the nuclei of the more normal spiral galaxies (M83, NGC 3627, NGC 3351) lie along the same relation, which provides some additional support for our choice to use the ULIRG CO-to-H$_2$ conversion factor for these regions. Nine pixels in the central regions of each of NGC 7469 and Circinus lie noticeably above the relation. These pixels contain emission from the bright AGN nucleus in each galaxy, and so the star formation rate surface density in these pixels has been overestimated. However, none of the other AGN nuclei (Table~\ref{table:physical}) show a clear offset from the Kennicutt-Schmidt relation. Color-coding  the points  by the CN/HCN intensity ratio shows no clear trend of this ratio along the Kennicutt-Schmidt relation. Although some large CN/HCN ratios are found towards the Circinus AGN nucleus, even larger values are found in some of the non-nuclear pixels in other galaxies.

Although the CN/HCN line ratio is remarkably constant within each
galaxy, the central regions of 4 of the 7 galaxies show small but statistically significant differences (Table~\ref{table:ratios}). 
With the highest spatial resolution in our sample (62 pc), Circinus shows a 35\% enhancement in the  CN/HCN line ratio in the central pixels impacted by its strong AGN. In contrast, NGC 7469 does not show an enhanced CN/HCN ratio near its luminous AGN, likely because the 300 pc resolution of the data mixes emission from the AGN and surrounding regions within the beam. Interestingly, the northern (starburst) nucleus of NGC 3256 shows a 15\% enhancement in the CN/HCN line ratio compared to its global average, while the southern (AGN) nucleus does not show an enhancement. The 470 pc resolution of the data may be diluting any enhanced line ratio from the southern AGN, while the northern starburst may cover a large enough area to produce a measurable increase in line ratio \citep[see also discussion in ][]{wilson2018}. Finally, NGC 1808 also shows an increase (20\%) in the CN/HCN ratio in its center; its relatively faint AGN combined with the extent of the area of the enhanced ratio (Fig.~\ref{fig:appendixfig3}) may imply that the enhancement is driven by the central starburst rather than the AGN.
The only galaxy to show a statistically significant decrease in its central CN/HCN ratio (of 20\%) is IRAS 13120-5453. At the emission peak, our data show that the ratio of the two CN fine structure line groupings is $\sim 1.8$, which implies a moderate optical depth of 0.4-0.5 for the brighter line \citep[see formula in ][]{tang2019A&A...629A...6T}. 
Thus, it seems likely that the CN optical depth is playing a role in reducing the CN brightness and hence the CN/HCN intensity ratio at the center of IRAS 13120-5253.



\subsection{Comparison with models}

What, if anything, can we infer about the physical properties such as the gas density and radiation field in these galaxies from the observed CN/HCN ratio? Our pixel scale ranges from 30 pc to 400 pc, probing scales from an entire molecular cloud to an ensemble of molecular clouds. This resolution implies that we are averaging over a complicated mixture of physical properties and geometries, which makes it very difficult to make direct comparisons between models and observations \citep{wolfire2022}. 

Protoplanetary disk models already demonstrate that the CN/HCN abundance and column density ratios have a complicated dependence on the local density and UV radiation field on small scales. \citet{chapillon2012} find that for densities of $5\times 10^4 - 10^6$ cm$^{-3}$, the disk-integrated CN/HCN column density ratio changes by a factor of 2-6 with a factor of 100 change in the UV intensity. \citet{bergner2021ApJS..257...11B} find that the CN/HCN column density ratio changes by a factor of 100 from regions with low (outer) to high (inner) HCN column density. However, the structured
density profiles assumed in these disk models make it difficult to
know how to apply these results to the more
complicated and less organized mix of densities seen on larger scales in
molecular clouds.

On the somewhat larger scales of cores and clouds, models often show more similarity between CN and HCN emission than would be expected from the plane-parallel PDR models of \citet{boger2005}. The PCA analysis of \citet{gratier2017} shows that all molecules correlate with H$_2$ column density, and also to a lesser degree with volume density; somewhat unexpectedly, HCN shows a strong positive correlation with UV illumination, second only to CCH and CN in strength. The time-dependent chemistry models of \citet{harada2019models} fit global molecular cloud spectra for W51 and M51 for many species (including HCN and CN) with a moderate density of $1-3 \times 10^3$ cm$^{-3}$, column densities $A_v \ge 4$ mag, and chemical ages of $10^5$ yr. They note that this density is lower than the critical density for most of the molecular species included in the analysis. \citet{harada2019models} find that the short chemical timescale used to fit the data is consistent with a typical turbulent mixing timescape to expose the gas to UV photons and reset the chemical clock. 
The range of physical resolutions (10-100 pc) used in \citet{harada2019models} is comparable to the range (30-400 pc) studied in our sample. Whether a similarly short turbulent mixing timescale applies to the high column density, high star formation rate regimes probed in our sample remains to be tested, although large line widths and short free-fall times seen in the most active galaxies \citep{wilson2019} imply a large degree of turbulence.

Given these complexities, it seems premature to attempt to infer typical gas densities or UV radiation field strengths from the observed CN/HCN  line ratios. The relatively constant CN/HCN ratio observed over 2 orders of magnitude in both  $\Sigma_{mol}$ and  $\Sigma_{SFR}$ suggests that CN and HCN trace similar gas. This constant line ratio in turn implies that CN is ``as good'' a dense gas tracer as HCN is  when averaged over 30-400 pc regions; we discuss this conclusion further in the next section. Further progress on clarifying if and how the CN/HCN ratio depends on density, UV radiation field, and their ratio is most likely to come from additional analyses of detailed data for individual Galactic molecular clouds, such as by applying the \citet{bron2018} analysis to molecular clouds from the LEGO survey \citep{kauffmann2017,barnes2020}.

\subsection{Implications of a (nearly) constant CN/HCN line ratio}

The strong correlation between CN $N=1-0$ and HCN $J=1-0$ shown in Figures~\ref{fig:ratio_vs_intensity} and ~\ref{fig:ratio_vs_ratio} suggests that, when averaged over scales of a few 100 pc, these two emission lines are tracing the same gas.
Figure~\ref{fig:ratio_vs_ratio} also shows that the CN/CO and HCN/CO line ratios track each other very well over a range of more than a factor of 10 in HCN/CO line ratio. The HCN/CO ratio is often used as a tracer of the dense gas fraction in galaxies, $f_{dense}$ \citep[e.g.,][]{gallagher2018,bemis2019,jimenez2019,beslic2021}. 
\citet{jimenez2019}
 present a detailed study of the HCN emission across 
9 nearby spiral galaxies (including NGC 3627). They find that the HCN/CO ratio (the fraction of dense gas) is positively correlated with the stellar
surface density and the molecular gas surface density. They find that the
infrared/HCN ratio (the SFR to dense gas ratio) is anti-correlated with the same 
two quantities. The HCN/CO ratio also varies strongly
in our sample (Figure~\ref{fig:ratio_vs_ratio}), extending to values 4 times larger than those seen in the analysis
of \citet{jimenez2019}, while the molecular gas surface density extends to a factor of 10 times higher values in our sample (Figure~\ref{fig:ratio_vs_intensity}). Further analysis of these line ratios with
star formation and molecular gas surface densities will be presented in 
a future paper.

The strong correlation between CN and HCN (Figures~\ref{fig:ratio_vs_intensity} and ~\ref{fig:ratio_vs_ratio}) suggests that the CN line will be  an equally good tracer of the dense gas in galaxies on scales of whole clouds and larger. One attractive aspect of using CN rather than HCN to trace the dense gas is that CN $N=1-0$ and CO $J=1-0$ can be (and often are) observed simultaneously with ALMA. Compared to the HCN/CO line ratio (which requires two separate observations), such simultaneous observations greatly reduce uncertainties in the CN/CO line ratio  from calibration  and the response function in the interferometric maps from the beam and $uv$ coverage.

An intriguing aspect of the observed correlation between the CN and HCN lines relates to their optical depths. The relative strength of the two main groupings of the CN $N=1-0$ transition can be used to derive optical depth \citep[e.g.,][]{tang2019A&A...629A...6T}, with a line ratio of 2 implying optically thin emission in both lines. The bulk of the CN emission is optically thin, e.g., in the starburst galaxy NGC253 \citep{meier2015} and in the W49 molecular cloud \citep{barnes2020}. In our sample, only the highest surface density region of IRAS~13120 shows some indication of higher CN optical depth, with an optical depth of 0.4-0.5 for the brighter line (see discussion in \S~\ref{subsec:constLineRatio}. On the other hand, HCN emission has been found to be optically thick in recent extragalactic studies \citep[][and references therein]{jimenezdonaire2017}, with optical depths in the range of 2-11 in the central regions. 

Assuming the gas producing the bulk of the HCN emission in galaxies is indeed optically thick, how does it arise that an optically thin line (CN) and an optically thick line (HCN) are so well correlated? We may gain insights from considering the comparison of the optically thick CO line and its optically thin isotopologue $^{13}$CO. Over scales from 10 pc up to kiloparsecs, the CO line correlates with the total molecular gas surface density quite well \citep[via the well-known CO-to-H$_2$ conversion factor, e.g., ][]{bolatto2013b}. This correlation traces back to the average physical properties of a collection of virialized giant molecular clouds \citep{dickman1986}. On similar scales, the optically thin $^{13}$CO line will also trace the total molecular gas surface density as long as its abundance relative to H$_2$ and its excitation temperature do not vary strongly from place to place or galaxy to galaxy. A  similar mechanism may be at play for HCN and CN, with HCN tracing a population of gravitationally bound or collapsing sub-clumps inside molecular clouds indirectly (as CO does for entire clouds) and CN tracing the same dense gas directly through its optically thin emission lines.

Finally, we return to our initial expectation when starting this project, namely that the CN emission would trace UV-irradiated dense gas while the HCN emission would trace colder, shielded gas. Such a distinct separation does not appear on the scales of 30-400 pc probed in our extragalactic sample. Using the star formation rate surface density as a proxy for the UV photon surface density, our data show only a very mild increase of the CN/HCN line ratio with the increasing star formation rate surface density (Figure~\ref{fig:ratio_vs_intensity}) or  star formation efficiency (Figure~\ref{fig:ratio_vs_ratio}). Studies of Galactic molecular clouds even find that the HCN emission is more spatially extended than the CN emission \citep{kauffmann2017,barnes2020}, which appears inconsistent with the slab-PDR models of \citet{boger2005}.

In this context, it would be very helpful to have  hydrodynamical simulations of CN, HCN, and CO in clumpy molecular clouds to compare to the wealth of observational datasets that are becoming available on these and other emission lines from Galactic molecular clouds \citep{kauffmann2017,bron2018,gratier2017,barnes2020}.
\citet{gaches2015} combined hydrodynamic simulations of a molecular cloud with a detailed astrochemical network to study how different molecular species trace different structures and density regimes. They predict that CN should trace the same gas as CO and have similarly high optical depth. However, this conclusion of high CN optical depth appears inconsistent with the relative strengths of the two primary fine-structure groupings of the CN $N=1-0$ line, which indicate that CN is optically thin \citep[e.g.,][]{meier2015}.
Thus, although the models of \citet{gaches2015} include both CN and HCN, it is difficult to draw any useful conclusions from these  models for a cloud-wide or extragalactic context.

\section{Conclusions}
\label{sec:concl}

We have used a sample of 9 nearby galaxies drawn from the ALMA archive to investigate the relationship between CN and HCN on scales ranging from 30 pc to 400 pc. The sample includes luminous and ultraluminous infrared galaxies, galaxies with a prominent AGN, galaxies with starburst nuclei, and spiral galaxies. We find the intensity ratio of the brighter fine-structure line grouping of the CN $N=1-0$ line relative to the HCN $J=1-0$ line to be remarkably constant (Figure~\ref{fig:ratio_vs_intensity}), with a mean value of $0.86 \pm 0.07$ (standard deviation 0.20). 
The constant ratio between the CN and HCN line intensities is somewhat unexpected, given that HCN is expected to reside in shielded, high column density regions while CN is expected to trace dense gas exposed to high ultraviolet radiation.
The CN/HCN line ratio shows no significant trend with molecular gas surface density, but does show a mild increase ($\sim 1.3$ per dex) with both increasing star formation rate surface density and star formation efficiency (the inverse of the molecular gas depletion time). Some AGN and starburst nuclei show small enhancements in their CN/HCN ratio, while other nuclei show no significant difference from their surrounding disks.

This nearly constant CN/HCN intensity ratio implies that CN, like HCN, may be useful as a tracer of dense, star-forming gas in nearby galaxies across a wide range of star formation rate surface densities. In particular, the CN/CO and HCN/CO ratios track each other very well (Figure~\ref{fig:ratio_vs_ratio}), which suggests that either ratio can be used as a measure of the dense gas fraction in galaxy disks. Since the CN $N=1-0$ line is often measured simultaneously with the  CO $J=1-0$ line when observing nearby galaxies with ALMA, this correlation between CN and HCN emission opens up the possibility of resolved studies of dense gas in large numbers of nearby galaxies using the ALMA archive.


\section*{Acknowledgements}

We thank the anonymous referee for detailed comments that significantly improved the content of this paper. We also thank Dr. Jiayi Sun for help in generating the electronic data table.
This paper makes use of the following ALMA data: \\
ADS/JAO.ALMA\#2011.0.00467.S, 
ADS/JAO.ALMA\#2011.0.00525.S, 
ADS/JAO.ALMA\#2011.0.00772.S, 
ADS/JAO.ALMA\#2012.1.00165.S, 
ADS/JAO.ALMA\#2012.1.01004.S, 
ADS/JAO.ALMA\#2013.1.00218.S, 
ADS/JAO.ALMA\#2013.1.00247.S, 
ADS/JAO.ALMA\#2013.1.00379.S, 
ADS/JAO.ALMA\#2013.1.00634.S, 
ADS/JAO.ALMA\#2013.1.00885.S, 
ADS/JAO.ALMA\#2013.1.00911.S, 
ADS/JAO.ALMA\#2013.1.01057.S, 
ADS/JAO.ALMA\#2015.1.00287.S, 
ADS/JAO.ALMA\#2015.1.00993.S, 
ADS/JAO.ALMA\#2015.1.01177.S, 
ADS/JAO.ALMA\#2015.1.01286.S, 
ADS/JAO.ALMA\#2015.1.01538.S. 
\\
ALMA is a partnership of ESO
(representing its member states), NSF (USA) and NINS (Japan), together
with NRC (Canada), MOST and ASIAA (Taiwan), and KASI (Republic of
Korea), in cooperation with the Republic of Chile. The Joint ALMA
Observatory is operated by ESO, AUI/NRAO and NAOJ. 
The National Radio
Astronomy Observatory is a facility of the National Science Foundation
operated under cooperative agreement by Associated Universities, Inc. 
%
%
%
This research has made use of the NASA/IPAC Extragalactic Database
(NED) which is operated by the Jet Propulsion Laboratory, California
Institute of Technology, under contract with the National Aeronautics
and Space Administration. 
This research has made use of the NASA/IPAC Extragalactic Database, which is funded by the National Aeronautics and Space Administration and operated by the California Institute of Technology. This research has also made use of  astropy, a community-developed core python package for astronomy \citep{astropy2013}.
CDW acknowledges financial support from the 
Canada Council for the Arts through a Killam Research Fellowship. The
research of CDW is supported by grants from the Natural Sciences and
Engineering Research Council of Canada and the Canada Research Chairs
program. CDW would like to thank NRAO for its hospitality
during the first year of this project; the computing resources available at
NRAO were essential for this project.
AB thanks the Ontario Trillium Scholarship for financial support during the completion of this work. BL acknowledges partial support from a NSERC Canada Graduate Scholarship-Doctoral and an Ontario Graduate Scholarship. OK acknowledges partial support from Natural Sciences and Engineering Research Council of Canada (NSERC) Undergraduate Summer Research Award. 

\section{Data availability}

The derived data generated in this research will be shared on reasonable request to the corresponding author.




\bibliographystyle{mnras}
\bibliography{killampaper1_v1} 

\begin{thebibliography}{}
\makeatletter
\relax
\def\mn@urlcharsother{\let\do\@makeother \do\$\do\&\do\#\do\^\do\_\do\%\do\~}
\def\mn@doi{\begingroup\mn@urlcharsother \@ifnextchar [ {\mn@doi@}
  {\mn@doi@[]}}
\def\mn@doi@[#1]#2{\def\@tempa{#1}\ifx\@tempa\@empty \href
  {http://dx.doi.org/#2} {doi:#2}\else \href {http://dx.doi.org/#2} {#1}\fi
  \endgroup}
\def\mn@eprint#1#2{\mn@eprint@#1:#2::\@nil}
\def\mn@eprint@arXiv#1{\href {http://arxiv.org/abs/#1} {{\tt arXiv:#1}}}
\def\mn@eprint@dblp#1{\href {http://dblp.uni-trier.de/rec/bibtex/#1.xml}
  {dblp:#1}}
\def\mn@eprint@#1:#2:#3:#4\@nil{\def\@tempa {#1}\def\@tempb {#2}\def\@tempc
  {#3}\ifx \@tempc \@empty \let \@tempc \@tempb \let \@tempb \@tempa \fi \ifx
  \@tempb \@empty \def\@tempb {arXiv}\fi \@ifundefined
  {mn@eprint@\@tempb}{\@tempb:\@tempc}{\expandafter \expandafter \csname
  mn@eprint@\@tempb\endcsname \expandafter{\@tempc}}}

\bibitem[\protect\citeauthoryear{{Aalto}, {Polatidis}, {H{\"u}ttemeister}  \&
  {Curran}}{{Aalto} et~al.}{2002}]{aalto2002}
{Aalto} S.,  {Polatidis} A.~G.,  {H{\"u}ttemeister} S.,   {Curran} S.~J.,
  2002, \mn@doi [\aap] {10.1051/0004-6361:20011514}, 381, 783

\bibitem[\protect\citeauthoryear{{Gao} \& {Solomon}}{{Gao} \&
  {Solomon}}{2004}]{gao2004}
{Gao} Y.,  {Solomon} P.~M.,  2004, \mn@doi [\apj] {10.1086/382999}, 606, 271

\bibitem[\protect\citeauthoryear{{Juneau}, {Narayanan}, {Moustakas}, {Shirley},
  {Bussmann}, {Kennicutt}  \& {Vanden Bout}}{{Juneau}
  et~al.}{2009}]{juneau2009}
{Juneau} S.,  {Narayanan} D.~T.,  {Moustakas} J.,  {Shirley} Y.~L.,  {Bussmann}
  R.~S.,  {Kennicutt} Jr. R.~C.,   {Vanden Bout} P.~A.,  2009, \mn@doi [\apj]
  {10.1088/0004-637X/707/2/1217}, 707, 1217

\bibitem[\protect\citeauthoryear{{Kennicutt} \& {Evans}}{{Kennicutt} \&
  {Evans}}{2012}]{kennicutt2012}
{Kennicutt} R.~C.,  {Evans} N.~J.,  2012, \mn@doi [\araa]
  {10.1146/annurev-astro-081811-125610}, 50, 531

\bibitem[\protect\citeauthoryear{{Wu}, {Evans}, {Shirley}  \& {Knez}}{{Wu}
  et~al.}{2010}]{wu2010}
{Wu} J.,  {Evans} II N.~J.,  {Shirley} Y.~L.,   {Knez} C.,  2010, \mn@doi
  [\apjs] {10.1088/0067-0049/188/2/313}, 188, 313

\makeatother
\end{thebibliography}


\begin{thebibliography}{}
\makeatletter
\relax
\def\mn@urlcharsother{\let\do\@makeother \do\$\do\&\do\#\do\^\do\_\do\%\do\~}
\def\mn@doi{\begingroup\mn@urlcharsother \@ifnextchar [ {\mn@doi@}
  {\mn@doi@[]}}
\def\mn@doi@[#1]#2{\def\@tempa{#1}\ifx\@tempa\@empty \href
  {http://dx.doi.org/#2} {doi:#2}\else \href {http://dx.doi.org/#2} {#1}\fi
  \endgroup}
\def\mn@eprint#1#2{\mn@eprint@#1:#2::\@nil}
\def\mn@eprint@arXiv#1{\href {http://arxiv.org/abs/#1} {{\tt arXiv:#1}}}
\def\mn@eprint@dblp#1{\href {http://dblp.uni-trier.de/rec/bibtex/#1.xml}
  {dblp:#1}}
\def\mn@eprint@#1:#2:#3:#4\@nil{\def\@tempa {#1}\def\@tempb {#2}\def\@tempc
  {#3}\ifx \@tempc \@empty \let \@tempc \@tempb \let \@tempb \@tempa \fi \ifx
  \@tempb \@empty \def\@tempb {arXiv}\fi \@ifundefined
  {mn@eprint@\@tempb}{\@tempb:\@tempc}{\expandafter \expandafter \csname
  mn@eprint@\@tempb\endcsname \expandafter{\@tempc}}}

\bibitem[\protect\citeauthoryear{{Arulanantham} et~al.,}{{Arulanantham}
  et~al.}{2020}]{arulanantham2020}
{Arulanantham} N.,  et~al., 2020, \mn@doi [\aj] {10.3847/1538-3881/ab789a},
  \href {https://ui.adsabs.harvard.edu/abs/2020AJ....159..168A} {159, 168}

\bibitem[\protect\citeauthoryear{{Astropy Collaboration} et~al.,}{{Astropy
  Collaboration} et~al.}{2013}]{astropy2013}
{Astropy Collaboration} et~al., 2013, \mn@doi [\aap]
  {10.1051/0004-6361/201322068}, \href
  {https://ui.adsabs.harvard.edu/abs/2013A&A...558A..33A} {558, A33}

\bibitem[\protect\citeauthoryear{{Barnes} et~al.,}{{Barnes}
  et~al.}{2020}]{barnes2020}
{Barnes} A.~T.,  et~al., 2020, \mn@doi [\mnras] {10.1093/mnras/staa1814}, \href
  {https://ui.adsabs.harvard.edu/abs/2020MNRAS.497.1972B} {497, 1972}

\bibitem[\protect\citeauthoryear{{Bemis} \& {Wilson}}{{Bemis} \&
  {Wilson}}{2019}]{bemis2019}
{Bemis} A.,  {Wilson} C.~D.,  2019, \mn@doi [\aj] {10.3847/1538-3881/ab041d},
  \href {https://ui.adsabs.harvard.edu/abs/2019AJ....157..131B} {157, 131}

\bibitem[\protect\citeauthoryear{{Bergner} et~al.,}{{Bergner}
  et~al.}{2021}]{bergner2021ApJS..257...11B}
{Bergner} J.~B.,  et~al., 2021, \mn@doi [\apjs] {10.3847/1538-4365/ac143a},
  \href {https://ui.adsabs.harvard.edu/abs/2021ApJS..257...11B} {257, 11}

\bibitem[\protect\citeauthoryear{{Be{\v{s}}li{\'c}} et~al.,}{{Be{\v{s}}li{\'c}}
  et~al.}{2021}]{beslic2021}
{Be{\v{s}}li{\'c}} I.,  et~al., 2021, \mn@doi [\mnras]
  {10.1093/mnras/stab1776}, \href
  {https://ui.adsabs.harvard.edu/abs/2021MNRAS.506..963B} {506, 963}

\bibitem[\protect\citeauthoryear{{Bigiel}, {Leroy}, {Walter}, {Brinks}, {de
  Blok}, {Madore}  \& {Thornley}}{{Bigiel} et~al.}{2008}]{bigiel2008}
{Bigiel} F.,  {Leroy} A.,  {Walter} F.,  {Brinks} E.,  {de Blok} W.~J.~G.,
  {Madore} B.,   {Thornley} M.~D.,  2008, \mn@doi [\aj]
  {10.1088/0004-6256/136/6/2846}, \href
  {https://ui.adsabs.harvard.edu/abs/2008AJ....136.2846B} {136, 2846}

\bibitem[\protect\citeauthoryear{{Boger} \& {Sternberg}}{{Boger} \&
  {Sternberg}}{2005}]{boger2005}
{Boger} G.~I.,  {Sternberg} A.,  2005, \mn@doi [\apj] {10.1086/432864}, 632,
  302

\bibitem[\protect\citeauthoryear{{Bolatto}, {Wolfire}  \& {Leroy}}{{Bolatto}
  et~al.}{2013}]{bolatto2013b}
{Bolatto} A.~D.,  {Wolfire} M.,   {Leroy} A.~K.,  2013, \mn@doi [\araa]
  {10.1146/annurev-astro-082812-140944}, 51, 207

\bibitem[\protect\citeauthoryear{{Brinkmann}, {Wyrowski}, {Kauffmann},
  {Colombo}, {Menten}, {Tang}  \& {G{\"u}sten}}{{Brinkmann}
  et~al.}{2020}]{brinkmann2020}
{Brinkmann} N.,  {Wyrowski} F.,  {Kauffmann} J.,  {Colombo} D.,  {Menten}
  K.~M.,  {Tang} X.~D.,   {G{\"u}sten} R.,  2020, \mn@doi [\aap]
  {10.1051/0004-6361/201936885}, \href
  {https://ui.adsabs.harvard.edu/abs/2020A&A...636A..39B} {636, A39}

\bibitem[\protect\citeauthoryear{{Bron} et~al.,}{{Bron}
  et~al.}{2018}]{bron2018}
{Bron} E.,  et~al., 2018, \mn@doi [\aap] {10.1051/0004-6361/201731833}, \href
  {https://ui.adsabs.harvard.edu/abs/2018A&A...610A..12B} {610, A12}

\bibitem[\protect\citeauthoryear{{Chapillon}, {Guilloteau}, {Dutrey},
  {Pi{\'e}tu}  \& {Gu{\'e}lin}}{{Chapillon} et~al.}{2012}]{chapillon2012}
{Chapillon} E.,  {Guilloteau} S.,  {Dutrey} A.,  {Pi{\'e}tu} V.,   {Gu{\'e}lin}
  M.,  2012, \mn@doi [\aap] {10.1051/0004-6361/201116762}, 537, A60

\bibitem[\protect\citeauthoryear{{Cicone}, {Maiolino}, {Aalto}, {Muller}  \&
  {Feruglio}}{{Cicone} et~al.}{2020}]{cicone2020A&A...633A.163C}
{Cicone} C.,  {Maiolino} R.,  {Aalto} S.,  {Muller} S.,   {Feruglio} C.,  2020,
  \mn@doi [\aap] {10.1051/0004-6361/201936800}, \href
  {https://ui.adsabs.harvard.edu/abs/2020A&A...633A.163C} {633, A163}

\bibitem[\protect\citeauthoryear{{Combes} et~al.,}{{Combes}
  et~al.}{2019}]{combes2019A&A...623A..79C}
{Combes} F.,  et~al., 2019, \mn@doi [\aap] {10.1051/0004-6361/201834560}, \href
  {https://ui.adsabs.harvard.edu/abs/2019A&A...623A..79C} {623, A79}

\bibitem[\protect\citeauthoryear{{Crosthwaite}, {Turner}, {Buchholz}, {Ho}  \&
  {Martin}}{{Crosthwaite} et~al.}{2002}]{crosthwaite2002AJ....123.1892C}
{Crosthwaite} L.~P.,  {Turner} J.~L.,  {Buchholz} L.,  {Ho} P. T.~P.,
  {Martin} R.~N.,  2002, \mn@doi [\aj] {10.1086/339479}, \href
  {https://ui.adsabs.harvard.edu/abs/2002AJ....123.1892C} {123, 1892}

\bibitem[\protect\citeauthoryear{{Cruz-Gonz{\'a}lez}
  et~al.,}{{Cruz-Gonz{\'a}lez} et~al.}{2020}]{cruz-gonzalez2020MNRAS.499.2042C}
{Cruz-Gonz{\'a}lez} I.,  et~al., 2020, \mn@doi [\mnras]
  {10.1093/mnras/staa2949}, \href
  {https://ui.adsabs.harvard.edu/abs/2020MNRAS.499.2042C} {499, 2042}

\bibitem[\protect\citeauthoryear{{Curran}, {Koribalski}  \& {Bains}}{{Curran}
  et~al.}{2008}]{curran2008MNRAS.389...63C}
{Curran} S.~J.,  {Koribalski} B.~S.,   {Bains} I.,  2008, \mn@doi [\mnras]
  {10.1111/j.1365-2966.2008.13574.x}, \href
  {https://ui.adsabs.harvard.edu/abs/2008MNRAS.389...63C} {389, 63}

\bibitem[\protect\citeauthoryear{{Dickman}, {Snell}  \& {Schloerb}}{{Dickman}
  et~al.}{1986}]{dickman1986}
{Dickman} R.~L.,  {Snell} R.~L.,   {Schloerb} F.~P.,  1986, \mn@doi [\apj]
  {10.1086/164604}, \href
  {https://ui.adsabs.harvard.edu/abs/1986ApJ...309..326D} {309, 326}

\bibitem[\protect\citeauthoryear{{Downes} \& {Solomon}}{{Downes} \&
  {Solomon}}{1998}]{downes1998}
{Downes} D.,  {Solomon} P.~M.,  1998, \mn@doi [\apj] {10.1086/306339}, 507, 615

\bibitem[\protect\citeauthoryear{{For}, {Koribalski}  \& {Jarrett}}{{For}
  et~al.}{2012}]{for2012MNRAS.425.1934F}
{For} B.~Q.,  {Koribalski} B.~S.,   {Jarrett} T.~H.,  2012, \mn@doi [\mnras]
  {10.1111/j.1365-2966.2012.21416.x}, \href
  {https://ui.adsabs.harvard.edu/abs/2012MNRAS.425.1934F} {425, 1934}

\bibitem[\protect\citeauthoryear{{Freedman} et~al.,}{{Freedman}
  et~al.}{2001}]{freedman2001}
{Freedman} W.~L.,  et~al., 2001, \mn@doi [\apj] {10.1086/320638}, \href
  {https://ui.adsabs.harvard.edu/abs/2001ApJ...553...47F} {553, 47}

\bibitem[\protect\citeauthoryear{{Gaches}, {Offner}, {Rosolowsky}  \&
  {Bisbas}}{{Gaches} et~al.}{2015}]{gaches2015}
{Gaches} B. A.~L.,  {Offner} S. S.~R.,  {Rosolowsky} E.~W.,   {Bisbas} T.~G.,
  2015, \mn@doi [\apj] {10.1088/0004-637X/799/2/235}, \href
  {https://ui.adsabs.harvard.edu/abs/2015ApJ...799..235G} {799, 235}

\bibitem[\protect\citeauthoryear{{Gallagher} et~al.,}{{Gallagher}
  et~al.}{2018}]{gallagher2018}
{Gallagher} M.~J.,  et~al., 2018, \mn@doi [\apj] {10.3847/1538-4357/aabad8},
  \href {https://ui.adsabs.harvard.edu/abs/2018ApJ...858...90G} {858, 90}

\bibitem[\protect\citeauthoryear{{Ganeshalingam}, {Li}  \&
  {Filippenko}}{{Ganeshalingam} et~al.}{2013}]{ganeshalingam2013}
{Ganeshalingam} M.,  {Li} W.,   {Filippenko} A.~V.,  2013, \mn@doi [\mnras]
  {10.1093/mnras/stt893}, \href
  {https://ui.adsabs.harvard.edu/abs/2013MNRAS.433.2240G} {433, 2240}

\bibitem[\protect\citeauthoryear{{Gao} \& {Solomon}}{{Gao} \&
  {Solomon}}{2004}]{gao2004}
{Gao} Y.,  {Solomon} P.~M.,  2004, \mn@doi [\apj] {10.1086/382999}, 606, 271

\bibitem[\protect\citeauthoryear{{Gratier} et~al.,}{{Gratier}
  et~al.}{2017}]{gratier2017}
{Gratier} P.,  et~al., 2017, \mn@doi [\aap] {10.1051/0004-6361/201629847},
  \href {http://adsabs.harvard.edu/abs/2017A%26A...599A.100G} {599, A100}

\bibitem[\protect\citeauthoryear{{Hacar}, {Bosman}  \& {van Dishoeck}}{{Hacar}
  et~al.}{2020}]{hacar2020}
{Hacar} A.,  {Bosman} A.~D.,   {van Dishoeck} E.~F.,  2020, \mn@doi [\aap]
  {10.1051/0004-6361/201936516}, \href
  {https://ui.adsabs.harvard.edu/abs/2020A&A...635A...4H} {635, A4}

\bibitem[\protect\citeauthoryear{{Harada}, {Sakamoto}, {Martin}, {Aalto},
  {Aladro}  \& {Sliwa}}{{Harada} et~al.}{2018}]{harada2018new}
{Harada} N.,  {Sakamoto} K.,  {Martin} S.,  {Aalto} S.,  {Aladro} R.,   {Sliwa}
  K.,  2018, \apj, 855, 49

\bibitem[\protect\citeauthoryear{{Harada}, {Nishimura}, {Watanabe}, {Yamamoto},
  {Aikawa}, {Sakai}  \& {Shimonishi}}{{Harada}
  et~al.}{2019a}]{harada2019models}
{Harada} N.,  {Nishimura} Y.,  {Watanabe} Y.,  {Yamamoto} S.,  {Aikawa} Y.,
  {Sakai} N.,   {Shimonishi} T.,  2019a, \mn@doi [\apj]
  {10.3847/1538-4357/aaf72a}, \href
  {https://ui.adsabs.harvard.edu/abs/2019ApJ...871..238H} {871, 238}

\bibitem[\protect\citeauthoryear{{Harada}, {Sakamoto}, {Mart{\'\i}n},
  {Watanabe}, {Aladro}, {Riquelme}  \& {Hirota}}{{Harada}
  et~al.}{2019b}]{harada2019m83}
{Harada} N.,  {Sakamoto} K.,  {Mart{\'\i}n} S.,  {Watanabe} Y.,  {Aladro} R.,
  {Riquelme} D.,   {Hirota} A.,  2019b, \mn@doi [\apj]
  {10.3847/1538-4357/ab41ff}, \href
  {https://ui.adsabs.harvard.edu/abs/2019ApJ...884..100H} {884, 100}

\bibitem[\protect\citeauthoryear{{Howell} et~al.,}{{Howell}
  et~al.}{2010}]{howell2010ApJ...715..572H}
{Howell} J.~H.,  et~al., 2010, \mn@doi [\apj] {10.1088/0004-637X/715/1/572},
  \href {https://ui.adsabs.harvard.edu/abs/2010ApJ...715..572H} {715, 572}

\bibitem[\protect\citeauthoryear{{Jim{\'e}nez-Donaire}
  et~al.,}{{Jim{\'e}nez-Donaire} et~al.}{2017}]{jimenezdonaire2017}
{Jim{\'e}nez-Donaire} M.~J.,  et~al., 2017, \mn@doi [\mnras]
  {10.1093/mnras/stw2996}, \href
  {https://ui.adsabs.harvard.edu/abs/2017MNRAS.466...49J} {466, 49}

\bibitem[\protect\citeauthoryear{{Jim{\'e}nez-Donaire}
  et~al.,}{{Jim{\'e}nez-Donaire} et~al.}{2019}]{jimenez2019}
{Jim{\'e}nez-Donaire} M.~J.,  et~al., 2019, \mn@doi [\apj]
  {10.3847/1538-4357/ab2b95}, \href
  {https://ui.adsabs.harvard.edu/abs/2019ApJ...880..127J} {880, 127}

\bibitem[\protect\citeauthoryear{{Kauffmann}, {Goldsmith}, {Melnick}, {Tolls},
  {Guzman}  \& {Menten}}{{Kauffmann} et~al.}{2017}]{kauffmann2017}
{Kauffmann} J.,  {Goldsmith} P.~F.,  {Melnick} G.,  {Tolls} V.,  {Guzman} A.,
  {Menten} K.~M.,  2017, \mn@doi [\aap] {10.1051/0004-6361/201731123}, \href
  {https://ui.adsabs.harvard.edu/abs/2017A&A...605L...5K} {605, L5}

\bibitem[\protect\citeauthoryear{{Kennicutt}}{{Kennicutt}}{1998}]{kennicutt1998}
{Kennicutt} Robert~C. J.,  1998, \mn@doi [\apj] {10.1086/305588}, \href
  {https://ui.adsabs.harvard.edu/abs/1998ApJ...498..541K} {498, 541}

\bibitem[\protect\citeauthoryear{{Kennicutt} \& {Evans}}{{Kennicutt} \&
  {Evans}}{2012}]{kennicutt2012}
{Kennicutt} R.~C.,  {Evans} N.~J.,  2012, \mn@doi [\araa]
  {10.1146/annurev-astro-081811-125610}, 50, 531

\bibitem[\protect\citeauthoryear{{Kennicutt} \& {de Los Reyes}}{{Kennicutt} \&
  {de Los Reyes}}{2021}]{kennicutt2021}
{Kennicutt} Robert~C. J.,  {de Los Reyes} M. A.~C.,  2021, \mn@doi [\apj]
  {10.3847/1538-4357/abd3a2}, \href
  {https://ui.adsabs.harvard.edu/abs/2021ApJ...908...61K} {908, 61}

\bibitem[\protect\citeauthoryear{{Lada}, {Depoy}, {Evans}  \& {Gatley}}{{Lada}
  et~al.}{1991}]{lada1991}
{Lada} E.~A.,  {Depoy} D.~L.,  {Evans} Neal~J. I.,   {Gatley} I.,  1991,
  \mn@doi [\apj] {10.1086/169881}, \href
  {https://ui.adsabs.harvard.edu/abs/1991ApJ...371..171L} {371, 171}

\bibitem[\protect\citeauthoryear{{Ledger}, {Wilson}, {Michiyama}, {Iono},
  {Aalto}, {Saito}, {Bemis}  \& {Aladro}}{{Ledger} et~al.}{2021}]{ledger2021}
{Ledger} B.,  {Wilson} C.~D.,  {Michiyama} T.,  {Iono} D.,  {Aalto} S.,
  {Saito} T.,  {Bemis} A.,   {Aladro} R.,  2021, \mn@doi [\mnras]
  {10.1093/mnras/stab1204}, \href
  {https://ui.adsabs.harvard.edu/abs/2021MNRAS.504.5863L} {504, 5863}

\bibitem[\protect\citeauthoryear{{Leroy} et~al.,}{{Leroy}
  et~al.}{2009}]{leroy2009AJ....137.4670L}
{Leroy} A.~K.,  et~al., 2009, \mn@doi [\aj] {10.1088/0004-6256/137/6/4670},
  \href {https://ui.adsabs.harvard.edu/abs/2009AJ....137.4670L} {137, 4670}

\bibitem[\protect\citeauthoryear{{Leroy} et~al.,}{{Leroy}
  et~al.}{2015}]{leroy2015}
{Leroy} A.~K.,  et~al., 2015, \mn@doi [\apj] {10.1088/0004-637X/801/1/25}, 801,
  25

\bibitem[\protect\citeauthoryear{{Leroy} et~al.,}{{Leroy}
  et~al.}{2017}]{leroy2017}
{Leroy} A.~K.,  et~al., 2017, \mn@doi [\apj] {10.3847/1538-4357/835/2/217},
  \href {https://ui.adsabs.harvard.edu/abs/2017ApJ...835..217L} {835, 217}

\bibitem[\protect\citeauthoryear{{Leroy} et~al.,}{{Leroy}
  et~al.}{2019}]{leroy2019ApJS..244...24L}
{Leroy} A.~K.,  et~al., 2019, \mn@doi [\apjs] {10.3847/1538-4365/ab3925}, \href
  {https://ui.adsabs.harvard.edu/abs/2019ApJS..244...24L} {244, 24}

\bibitem[\protect\citeauthoryear{{Leroy} et~al.,}{{Leroy}
  et~al.}{2021}]{leroy2021ApJS..257...43L}
{Leroy} A.~K.,  et~al., 2021, \mn@doi [\apjs] {10.3847/1538-4365/ac17f3}, \href
  {https://ui.adsabs.harvard.edu/abs/2021ApJS..257...43L} {257, 43}

\bibitem[\protect\citeauthoryear{{Levrier}, {Le Petit}, {Hennebelle},
  {Lesaffre}, {Gerin}  \& {Falgarone}}{{Levrier}
  et~al.}{2012}]{levrier2012A&A...544A..22L}
{Levrier} F.,  {Le Petit} F.,  {Hennebelle} P.,  {Lesaffre} P.,  {Gerin} M.,
  {Falgarone} E.,  2012, \mn@doi [\aap] {10.1051/0004-6361/201218865}, \href
  {https://ui.adsabs.harvard.edu/abs/2012A&A...544A..22L} {544, A22}

\bibitem[\protect\citeauthoryear{{Lin} et~al.,}{{Lin}
  et~al.}{2019}]{lin2019ApJ...884L..33L}
{Lin} L.,  et~al., 2019, \mn@doi [\apjl] {10.3847/2041-8213/ab4815}, \href
  {https://ui.adsabs.harvard.edu/abs/2019ApJ...884L..33L} {884, L33}

\bibitem[\protect\citeauthoryear{{Mart{\'\i}n} et~al.,}{{Mart{\'\i}n}
  et~al.}{2021}]{martin2021A&A...656A..46M}
{Mart{\'\i}n} S.,  et~al., 2021, \mn@doi [\aap] {10.1051/0004-6361/202141567},
  \href {https://ui.adsabs.harvard.edu/abs/2021A&A...656A..46M} {656, A46}

\bibitem[\protect\citeauthoryear{{McMullin}, {Waters}, {Schiebel}, {Young}  \&
  {Golap}}{{McMullin} et~al.}{2007}]{mcmullin2007}
{McMullin} J.~P.,  {Waters} B.,  {Schiebel} D.,  {Young} W.,   {Golap} K.,
  2007, in {Shaw} R.~A.,  {Hill} F.,   {Bell} D.~J.,  eds,  Astronomical
  Society of the Pacific Conference Series Vol. 376, Astronomical Data Analysis
  Software and Systems XVI. p.~127

\bibitem[\protect\citeauthoryear{{Meier} et~al.,}{{Meier}
  et~al.}{2015}]{meier2015}
{Meier} D.~S.,  et~al., 2015, \mn@doi [\apj] {10.1088/0004-637X/801/1/63}, 801,
  63

\bibitem[\protect\citeauthoryear{{Meijerink} \& {Spaans}}{{Meijerink} \&
  {Spaans}}{2005}]{meijerink2005}
{Meijerink} R.,  {Spaans} M.,  2005, \mn@doi [\aap]
  {10.1051/0004-6361:20042398}, 436, 397

\bibitem[\protect\citeauthoryear{{Meijerink}, {Spaans}  \&
  {Israel}}{{Meijerink} et~al.}{2007}]{meijerink2007}
{Meijerink} R.,  {Spaans} M.,   {Israel} F.~P.,  2007, \mn@doi [\aap]
  {10.1051/0004-6361:20066130}, 461, 793

\bibitem[\protect\citeauthoryear{{Murphy} et~al.,}{{Murphy}
  et~al.}{2011}]{murphy2011}
{Murphy} E.~J.,  et~al., 2011, \mn@doi [\apj] {10.1088/0004-637X/737/2/67},
  \href {https://ui.adsabs.harvard.edu/abs/2011ApJ...737...67M} {737, 67}

\bibitem[\protect\citeauthoryear{{Nakajima}, {Takano}, {Kohno}, {Harada}  \&
  {Herbst}}{{Nakajima} et~al.}{2018}]{nakajima2018}
{Nakajima} T.,  {Takano} S.,  {Kohno} K.,  {Harada} N.,   {Herbst} E.,  2018,
  \mn@doi [\pasj] {10.1093/pasj/psx153}, \href
  {https://ui.adsabs.harvard.edu/abs/2018PASJ...70....7N} {70, 7}

\bibitem[\protect\citeauthoryear{{Papadopoulos}, {van der Werf}, {Xilouris},
  {Isaak}, {Gao}  \& {M{\"u}hle}}{{Papadopoulos}
  et~al.}{2012}]{papadopoulos2012MNRAS.426.2601P}
{Papadopoulos} P.~P.,  {van der Werf} P.~P.,  {Xilouris} E.~M.,  {Isaak} K.~G.,
   {Gao} Y.,   {M{\"u}hle} S.,  2012, \mn@doi [\mnras]
  {10.1111/j.1365-2966.2012.21001.x}, \href
  {https://ui.adsabs.harvard.edu/abs/2012MNRAS.426.2601P} {426, 2601}

\bibitem[\protect\citeauthoryear{{Radburn-Smith} et~al.,}{{Radburn-Smith}
  et~al.}{2011}]{radburn2011}
{Radburn-Smith} D.~J.,  et~al., 2011, \mn@doi [\apjs]
  {10.1088/0067-0049/195/2/18}, \href
  {https://ui.adsabs.harvard.edu/abs/2011ApJS..195...18R} {195, 18}

\bibitem[\protect\citeauthoryear{{Saha}, {Thim}, {Tammann}, {Reindl}  \&
  {Sandage}}{{Saha} et~al.}{2006}]{saha2006}
{Saha} A.,  {Thim} F.,  {Tammann} G.~A.,  {Reindl} B.,   {Sandage} A.,  2006,
  \mn@doi [\apjs] {10.1086/503800}, 165, 108

\bibitem[\protect\citeauthoryear{{Saito} et~al.,}{{Saito}
  et~al.}{2022}]{saito2022arXiv220706448S}
{Saito} T.,  et~al., 2022, arXiv e-prints, \href
  {https://ui.adsabs.harvard.edu/abs/2022arXiv220706448S} {p. arXiv:2207.06448}

\bibitem[\protect\citeauthoryear{{Sakamoto}, {Aalto}, {Combes}, {Evans}  \&
  {Peck}}{{Sakamoto} et~al.}{2014}]{sakamoto2014}
{Sakamoto} K.,  {Aalto} S.,  {Combes} F.,  {Evans} A.,   {Peck} A.,  2014,
  \mn@doi [\apj] {10.1088/0004-637X/797/2/90}, 797, 90

\bibitem[\protect\citeauthoryear{{Sakamoto} et~al.,}{{Sakamoto}
  et~al.}{2017}]{sakamoto2017}
{Sakamoto} K.,  et~al., 2017, \mn@doi [\apj] {10.3847/1538-4357/aa8f4b}, \href
  {https://ui.adsabs.harvard.edu/abs/2017ApJ...849...14S} {849, 14}

\bibitem[\protect\citeauthoryear{{Salak}, {Nakai}  \& {Kitamoto}}{{Salak}
  et~al.}{2014}]{salak2014PASJ...66...96S}
{Salak} D.,  {Nakai} N.,   {Kitamoto} S.,  2014, \mn@doi [\pasj]
  {10.1093/pasj/psu074}, \href
  {https://ui.adsabs.harvard.edu/abs/2014PASJ...66...96S} {66, 96}

\bibitem[\protect\citeauthoryear{{Sanders}, {Mazzarella}, {Kim}, {Surace}  \&
  {Soifer}}{{Sanders} et~al.}{2003}]{sanders2003}
{Sanders} D.~B.,  {Mazzarella} J.~M.,  {Kim} D.-C.,  {Surace} J.~A.,   {Soifer}
  B.~T.,  2003, \mn@doi [\aj] {10.1086/376841}, 126, 1607

\bibitem[\protect\citeauthoryear{{Sandstrom} et~al.,}{{Sandstrom}
  et~al.}{2013}]{sandstrom2013ApJ...777....5S}
{Sandstrom} K.~M.,  et~al., 2013, \mn@doi [\apj] {10.1088/0004-637X/777/1/5},
  \href {https://ui.adsabs.harvard.edu/abs/2013ApJ...777....5S} {777, 5}

\bibitem[\protect\citeauthoryear{{Schruba} et~al.,}{{Schruba}
  et~al.}{2011}]{schruba2011}
{Schruba} A.,  et~al., 2011, \mn@doi [\aj] {10.1088/0004-6256/142/2/37}, \href
  {https://ui.adsabs.harvard.edu/abs/2011AJ....142...37S} {142, 37}

\bibitem[\protect\citeauthoryear{{Scoville} et~al.,}{{Scoville}
  et~al.}{2017}]{scoville2017}
{Scoville} N.,  et~al., 2017, \mn@doi [\apj] {10.3847/1538-4357/836/1/66},
  \href {https://ui.adsabs.harvard.edu/abs/2017ApJ...836...66S} {836, 66}

\bibitem[\protect\citeauthoryear{{Shao}, {Koribalski}, {Wang}, {Ho}  \&
  {Staveley-Smith}}{{Shao} et~al.}{2018}]{shao2018}
{Shao} L.,  {Koribalski} B.~S.,  {Wang} J.,  {Ho} L.~C.,   {Staveley-Smith} L.,
   2018, \mn@doi [\mnras] {10.1093/mnras/sty1608}, \href
  {https://ui.adsabs.harvard.edu/abs/2018MNRAS.479.3509S} {479, 3509}

\bibitem[\protect\citeauthoryear{{Shimajiri} et~al.,}{{Shimajiri}
  et~al.}{2017}]{shimajiri2017}
{Shimajiri} Y.,  et~al., 2017, \mn@doi [\aap] {10.1051/0004-6361/201730633},
  \href {https://ui.adsabs.harvard.edu/abs/2017A&A...604A..74S} {604, A74}

\bibitem[\protect\citeauthoryear{{Shirley}}{{Shirley}}{2015}]{shirley2015}
{Shirley} Y.~L.,  2015, \mn@doi [\pasp] {10.1086/680342}, \href
  {https://ui.adsabs.harvard.edu/abs/2015PASP..127..299S} {127, 299}

\bibitem[\protect\citeauthoryear{{Sliwa}, {Wilson}, {Aalto}  \&
  {Privon}}{{Sliwa} et~al.}{2017}]{sliwa2017ApJ...840L..11S}
{Sliwa} K.,  {Wilson} C.~D.,  {Aalto} S.,   {Privon} G.~C.,  2017, \mn@doi
  [\apjl] {10.3847/2041-8213/aa6ea4}, \href
  {https://ui.adsabs.harvard.edu/abs/2017ApJ...840L..11S} {840, L11}

\bibitem[\protect\citeauthoryear{{Sorce}, {Tully}, {Courtois}, {Jarrett},
  {Neill}  \& {Shaya}}{{Sorce} et~al.}{2014}]{sorce2014}
{Sorce} J.~G.,  {Tully} R.~B.,  {Courtois} H.~M.,  {Jarrett} T.~H.,  {Neill}
  J.~D.,   {Shaya} E.~J.,  2014, \mn@doi [\mnras] {10.1093/mnras/stu1450},
  \href {https://ui.adsabs.harvard.edu/abs/2014MNRAS.444..527S} {444, 527}

\bibitem[\protect\citeauthoryear{{Sun} et~al.,}{{Sun} et~al.}{2018}]{sun2018}
{Sun} J.,  et~al., 2018, \mn@doi [\apj] {10.3847/1538-4357/aac326}, \href
  {https://ui.adsabs.harvard.edu/abs/2018ApJ...860..172S} {860, 172}

\bibitem[\protect\citeauthoryear{{Tacconi} et~al.,}{{Tacconi}
  et~al.}{2013}]{tacconi2013}
{Tacconi} L.~J.,  et~al., 2013, \mn@doi [\apj] {10.1088/0004-637X/768/1/74},
  \href {https://ui.adsabs.harvard.edu/abs/2013ApJ...768...74T} {768, 74}

\bibitem[\protect\citeauthoryear{{Takano}, {Nakajima}  \& {Kohno}}{{Takano}
  et~al.}{2019}]{takano2019PASJ...71S..20T}
{Takano} S.,  {Nakajima} T.,   {Kohno} K.,  2019, \mn@doi [\pasj]
  {10.1093/pasj/psz020}, \href
  {https://ui.adsabs.harvard.edu/abs/2019PASJ...71S..20T} {71, S20}

\bibitem[\protect\citeauthoryear{{Tang} et~al.,}{{Tang}
  et~al.}{2019}]{tang2019A&A...629A...6T}
{Tang} X.~D.,  et~al., 2019, \mn@doi [\aap] {10.1051/0004-6361/201935603},
  \href {https://ui.adsabs.harvard.edu/abs/2019A&A...629A...6T} {629, A6}

\bibitem[\protect\citeauthoryear{{Tully}, {Rizzi}, {Shaya}, {Courtois},
  {Makarov}  \& {Jacobs}}{{Tully} et~al.}{2009}]{tully2009}
{Tully} R.~B.,  {Rizzi} L.,  {Shaya} E.~J.,  {Courtois} H.~M.,  {Makarov}
  D.~I.,   {Jacobs} B.~A.,  2009, \mn@doi [\aj] {10.1088/0004-6256/138/2/323},
  \href {https://ui.adsabs.harvard.edu/abs/2009AJ....138..323T} {138, 323}

\bibitem[\protect\citeauthoryear{{Ueda} et~al.,}{{Ueda}
  et~al.}{2017}]{ueda2017PASJ...69....6U}
{Ueda} J.,  et~al., 2017, \mn@doi [\pasj] {10.1093/pasj/psw110}, \href
  {https://ui.adsabs.harvard.edu/abs/2017PASJ...69....6U} {69, 6}

\bibitem[\protect\citeauthoryear{{Ueda} et~al.,}{{Ueda}
  et~al.}{2021}]{ueda2021ApJS..257...57U}
{Ueda} J.,  et~al., 2021, \mn@doi [\apjs] {10.3847/1538-4365/ac257a}, \href
  {https://ui.adsabs.harvard.edu/abs/2021ApJS..257...57U} {257, 57}

\bibitem[\protect\citeauthoryear{{Uematsu}, {Ueda}, {Tanimoto}, {Kawamuro},
  {Setoguchi}, {Ogawa}, {Yamada}  \& {Odaka}}{{Uematsu}
  et~al.}{2021}]{uematsu2021ApJ...913...17U}
{Uematsu} R.,  {Ueda} Y.,  {Tanimoto} A.,  {Kawamuro} T.,  {Setoguchi} K.,
  {Ogawa} S.,  {Yamada} S.,   {Odaka} H.,  2021, \mn@doi [\apj]
  {10.3847/1538-4357/abf0a2}, \href
  {https://ui.adsabs.harvard.edu/abs/2021ApJ...913...17U} {913, 17}

\bibitem[\protect\citeauthoryear{{Usero} et~al.,}{{Usero}
  et~al.}{2015}]{usero2015}
{Usero} A.,  et~al., 2015, \mn@doi [\aj] {10.1088/0004-6256/150/4/115}, \href
  {https://ui.adsabs.harvard.edu/abs/2015AJ....150..115U} {150, 115}

\bibitem[\protect\citeauthoryear{{Walter} et~al.,}{{Walter}
  et~al.}{2017}]{walter2017}
{Walter} F.,  et~al., 2017, \mn@doi [\apj] {10.3847/1538-4357/835/2/265}, 835,
  265

\bibitem[\protect\citeauthoryear{{Watanabe}, {Nishimura}, {Harada}, {Sakai},
  {Shimonishi}, {Aikawa}, {Kawamura}  \& {Yamamoto}}{{Watanabe}
  et~al.}{2017}]{watanabe2017}
{Watanabe} Y.,  {Nishimura} Y.,  {Harada} N.,  {Sakai} N.,  {Shimonishi} T.,
  {Aikawa} Y.,  {Kawamura} A.,   {Yamamoto} S.,  2017, \mn@doi [\apj]
  {10.3847/1538-4357/aa7ece}, 845, 116

\bibitem[\protect\citeauthoryear{{Wilson}}{{Wilson}}{2018}]{wilson2018}
{Wilson} C.~D.,  2018, \mn@doi [\mnras] {10.1093/mnras/sty845}, \href
  {https://ui.adsabs.harvard.edu/abs/2018MNRAS.477.2926W} {477, 2926}

\bibitem[\protect\citeauthoryear{{Wilson}, {Elmegreen}, {Bemis}  \&
  {Brunetti}}{{Wilson} et~al.}{2019}]{wilson2019}
{Wilson} C.~D.,  {Elmegreen} B.~G.,  {Bemis} A.,   {Brunetti} N.,  2019,
  \mn@doi [\apj] {10.3847/1538-4357/ab31f3}, \href
  {https://ui.adsabs.harvard.edu/abs/2019ApJ...882....5W} {882, 5}

\bibitem[\protect\citeauthoryear{{Wolfire}, {Vallini}  \& {Chevance}}{{Wolfire}
  et~al.}{2022}]{wolfire2022}
{Wolfire} M.~G.,  {Vallini} L.,   {Chevance} M.,  2022, arXiv e-prints, \href
  {https://ui.adsabs.harvard.edu/abs/2022arXiv220205867W} {p. arXiv:2202.05867}

\bibitem[\protect\citeauthoryear{{Wu}, {Evans}, {Gao}, {Solomon}, {Shirley}  \&
  {Vanden Bout}}{{Wu} et~al.}{2005}]{wu2005}
{Wu} J.,  {Evans} Neal~J. I.,  {Gao} Y.,  {Solomon} P.~M.,  {Shirley} Y.~L.,
  {Vanden Bout} P.~A.,  2005, \mn@doi [\apjl] {10.1086/499623}, \href
  {https://ui.adsabs.harvard.edu/abs/2005ApJ...635L.173W} {635, L173}

\bibitem[\protect\citeauthoryear{{Yamada}, {Ueda}, {Tanimoto}, {Imanishi},
  {Toba}, {Ricci}  \& {Privon}}{{Yamada}
  et~al.}{2021}]{yamada2021ApJS..257...61Y}
{Yamada} S.,  {Ueda} Y.,  {Tanimoto} A.,  {Imanishi} M.,  {Toba} Y.,  {Ricci}
  C.,   {Privon} G.~C.,  2021, \mn@doi [\apjs] {10.3847/1538-4365/ac17f5},
  \href {https://ui.adsabs.harvard.edu/abs/2021ApJS..257...61Y} {257, 61}

\bibitem[\protect\citeauthoryear{{Yamashita} et~al.,}{{Yamashita}
  et~al.}{2017}]{yamashita2017ApJ...844...96Y}
{Yamashita} T.,  et~al., 2017, \mn@doi [\apj] {10.3847/1538-4357/aa7af1}, \href
  {https://ui.adsabs.harvard.edu/abs/2017ApJ...844...96Y} {844, 96}

\bibitem[\protect\citeauthoryear{{Young}, {Meier}, {Bureau}, {Crocker}, {Davis}
   \& {Topal}}{{Young} et~al.}{2021}]{young2021ApJ...909...98Y}
{Young} L.~M.,  {Meier} D.~S.,  {Bureau} M.,  {Crocker} A.,  {Davis} T.~A.,
  {Topal} S.,  2021, \mn@doi [\apj] {10.3847/1538-4357/abe126}, \href
  {https://ui.adsabs.harvard.edu/abs/2021ApJ...909...98Y} {909, 98}

\bibitem[\protect\citeauthoryear{{Young}, {Meier}, {Crocker}, {Davis}  \&
  {Topal}}{{Young} et~al.}{2022}]{young2022ApJ...933...90Y}
{Young} L.~M.,  {Meier} D.~S.,  {Crocker} A.,  {Davis} T.~A.,   {Topal} S.,
  2022, \mn@doi [\apj] {10.3847/1538-4357/ac7149}, \href
  {https://ui.adsabs.harvard.edu/abs/2022ApJ...933...90Y} {933, 90}

\bibitem[\protect\citeauthoryear{{de los Reyes} \& {Kennicutt}}{{de los Reyes}
  \& {Kennicutt}}{2019}]{delosreyes2019}
{de los Reyes} M. A.~C.,  {Kennicutt} Robert~C. J.,  2019, \mn@doi [\apj]
  {10.3847/1538-4357/aafa82}, \href
  {https://ui.adsabs.harvard.edu/abs/2019ApJ...872...16D} {872, 16}

\makeatother
\end{thebibliography}




\appendix

\section{Binned data table}

\begin{table*}
	\centering
	\caption{Binned data used in this paper}
	\label{table:pixel_data}
	\begin{tabular}{lccccccccccccc} 
		\hline
Galaxy\footnotemark[1] & R.A. & Decl. 
& $\log \Sigma_{\rm mol}$ & $\sigma({\log \Sigma_{\rm mol}})$ 
& $\log \Sigma_{\rm SFR}$ & $\sigma({\log \Sigma_{\rm SFR}})$ & $\log {\rm SFE}$\footnotemark[2] 
\\
		\hline
iras13120\_0 & 198.77642 & -55.15669 & 3.0018 & 0.0218 & 1.2239 & 0.0596 & -7.7779
 \\
ngc3256\_0 & 156.96323 & -43.90619 & 2.7235 & 0.0217 & 0.1740 & 0.0996 & -8.5495
 \\ 
vv114\_0 & 16.94853 & -17.50732 & 2.3491& 0.0218 & 0.5215 & 0.0490 & -7.8276 \\ 
ngc7469\_0 & 345.81557 & 8.8734 & 2.6564 & 0.0222 & 0.3385 & 0.1093 & -8.3179 \\
ngc1808\_0  & 76.928 & -37.51555 & 2.2135 & 0.0220 & 0.0055 & 0.0252 & -8.2081  \\
m83\_0 & 204.25383 & -29.86865 & 2.8286 & 0.0218 & 0.2858 & 0.0933 & -8.5429 \\ 
circinus\_0 & 213.28961 & -65.34291 & 2.5551 & 0.0218 & 0.0680 & 0.0927 & -8.4871 \\
ngc3627\_0 & 170.06983 & 12.97819 & 2.2200 & 0.0217 & 0.0453 & 0.0365 & -8.1747  \\
ngc3351\_0 & 160.99055 & 11.70128 & 2.1935 & 0.0221 & -0.5528 & 0.0669 & -8.7462  \\
		\hline
Galaxy & $\sigma({\log \rm SFE})$\footnotemark[2]
& $\log {\rm CN/HCN}$ & $\sigma({\log \rm CN/HCN})$ 
& $\log {\rm CN/CO}$ & $\sigma({\log \rm CN/CO})$ 
& $\log {\rm HCN/CO}$ & $\sigma({\log \rm HCN/CO})$ \\
		\hline
iras13120\_0 & 0.0635 & 0.2966 & 0.0503 & -0.8669 & 0.0345 & -1.1645 & 0.0479 \\
ngc3256\_0 & 0.1020 & -0.1331 & 0.0321 & -1.4656 & 0.0317 & -1.3335 & 0.0311 \\ 
vv114\_0 & 0.0537 & 0.3411 & 0.0420 & -1.1773 & 0.0333 & -1.5195 & 0.0401 \\ 
ngc7469\_0 & 0.1116 & -0.3261 & 0.0930 & -1.5069 & 0.0850 & -1.1818 & 0.0491 \\
ngc1808\_0  & 0.0335 & -0.3540 & 0.0458 & -1.5845 & 0.0451 & -1.2315 & 0.0322  \\
m83\_0 & 0.0958 & -0.3400 & 0.0447 & -1.4611 & 0.0444 & -1.1220 & 0.0312  \\ 
circinus\_0 & 0.0952 & -0.0624 & 0.0370 & -1.5148 & 0.0365 & -1.4534 & 0.0313  \\
ngc3627\_0 & 0.0425 & -0.0214 & 0.0532 & -1.5705 & 0.0415 & -1.5501 & 0.0452 \\
ngc3351\_0 & 0.0704 & -0.2957 & 0.0416 & -1.4719 & 0.0395 & -1.1771 & 0.0338 \\
		\hline
	\end{tabular}
\\
\footnotemark[1] This table is available in its entirety in machine-readable form. \\
\footnotemark[2] SFE = $\Sigma_{\rm SFR}/\Sigma_{\rm mol}$
\end{table*}

We provide the binned data used in this paper as a machine-readable table (Table~\ref{table:pixel_data}). In this table, each row reports our measurements for one pixel in one galaxy. The contents of the rows are as follows:
\begin{enumerate}
    \item The name of the galaxy tagged with a pixel identifier, and the central coordinates of the pixel in decimal degrees.
    \item The logarithmic value and logarithmic uncertainty for: molecular gas surface density, $\Sigma_{\rm mol}$; the star formation rate surface density, $\Sigma_{\rm SFR}$; the star formation efficiency, SFE (equal to  the inverse of the molecular gas depletion time); the CN/HCN intensity ratio; the CN/CO intensity ratio; and the HCN/CO intensity ratio. All intensity ratios are calculated from images with units of K km s$^{-1}$.
\end{enumerate}

\section{Formal uncertainties on moment 1 and moment 2 maps}\label{app:uncerts}

We calculate the moment 1 map, which is the intensity-weighted velocity field $\bar{v}$, of a molecular transition via

\begin{equation} \label{eq:mom1}
    \bar{v} = \frac{\sum T_i v_i}{\sum T_i}
\end{equation}

\noindent where $T_i$ is the intensity in channel $i$ and $v_i$ is the velocity at channel $i$. The uncertainty on $T_i$ is assumed to be uniform across the cube and is the rms uncertainty $\sigma_T$. We then calculate the moment 2 or observed velocity dispersion map, $\sigma_v \equiv \textsc{mom}_2$, of the molecular line via

\begin{equation}
    \textsc{mom}_2 = \sqrt{\frac{\sum T_i (v_i - \bar{v})^2}{\sum T_i}}
\end{equation}

\noindent where $\bar{v}$ is the mean velocity determined via Eqn. \ref{eq:mom1}.
In this derivation, we use the notation {MOM}$_2$ instead of the more common notation $\sigma_v$ for the measured values in the moment 2 map, so as to avoid confusion with the derived uncertainties, which are also commonly denoted by $\sigma_x$.

\subsection{Uncertainty on the intensity weighted velocity, $\bar{v}$ (moment 1)}

We first calculate the uncertainty on the velocity field, $\sigma_{\bar{v}}$. We define the function $f = \sum T_i (v_i - v_\mathrm{mid})$, where $v_\mathrm{mid} = \sum v_i /N$, and $N$ is the number of channels summed to calculate $\bar{v}$. This variable $f$ is related to $\bar{v}$ via

\begin{align}
    \frac{f}{\sum T_i} &= \frac{\sum T_i v_i}{\sum T_i} - v_\mathrm{mid} \\
    &= \bar{v} - v_\mathrm{mid} \\
    \rightarrow \bar{v} &= \frac{f}{\sum T_i} + v_\mathrm{mid}
\end{align}

\noindent We assume the uncertainty on $v_\mathrm{mid}$ to be zero, and neglect any uncertainty from the division of $\sum T_i$. The uncertainty on $\bar{v}$, $\sigma_{\bar{v}}$, is then 

\begin{equation} \label{eq:sigv_sigv}
    \sigma_{\bar{v}} = \frac{{\sigma_f}}{\sum {T_i}}
\end{equation}

\noindent The uncertainty on $f$ is

\begin{equation}
    \sigma_f = \sqrt{ \sum_i^N \left[\sigma_T (v_i - v_\mathrm{mid})\right]^2}
\end{equation}

\noindent In the case where $N$ is an odd number, we can show that

\begin{equation}
     (v_i - v_\mathrm{mid}) = \begin{cases}
    0 & \text{for } N=1 \\
    \pm \Delta v &  \text{for } N=3 \\
    \pm 2\Delta v &  \text{for } N=5 \\
    0, \pm \Delta v, \pm 2\Delta v, ... \pm\frac{1}{2}(N-1)\Delta v & \text{for } N
    \end{cases}
\end{equation}

\noindent where $\Delta v$ is the channel width (velocity resolution) of our cube, assumed to be constant.  The uncertainty then simplifies to

\begin{align}
    \sigma_f &= \sigma_T\Delta v  \left(2\sum_i^{\frac{N-1}{2}} k^2\right)^{1/2}
\end{align}

\noindent We then use the relation $\sum_i^n k^2 = n(n+1)(2n+1)/6$, where $n=(N-1)/2$ to obtain

\begin{align}
    \sigma_f &= \sigma_T\Delta v \sqrt{\left(2\frac{n(n+1)(2n+1)}{6}\right)} \\
    &= \sigma_T\Delta v \sqrt{\frac{1}{3}\left[\left(\frac{N-1}{2}\right)\left(\frac{N+1}{2}\right)N \right]} \\
    &=  \frac{\sigma_T\Delta v}{2\sqrt{3}}\sqrt{N\left({N^2-1}\right) } \\
    & \approx \frac{\sigma_T\Delta v}{2\sqrt{3}} N\sqrt{N}
\end{align}

\noindent in the limit of large $N$. Plugging $\sigma_f$ back into Eq. \ref{eq:sigv_sigv} gives

\begin{align} 
    \sigma_{\bar{v}} &\approx \frac{1}{\sum {T_i}} \frac{\sigma_T\Delta v}{2\sqrt{3}} N\sqrt{N} \\
    &= \frac{\Delta_\textsc{BW}}{2\sqrt{3}} \frac{\sigma_I}{I} 
\end{align}

\noindent where we have inserted the integrated intensity $I =\sum {T_i} \Delta v $, the uncertainty on the integrated intensity $\sigma_I = \sqrt{N} \sigma_T \Delta v$, and the full line width over which $\bar{v}$ and $\sigma_{\bar{v}}$ are calculated, $\Delta V_{line} = N \Delta v$.

\subsection{Uncertainty on the observed velocity dispersion, {MOM}$_2$ 
(moment 2)}

We first calculate the uncertainty on the variance, $\textsc{var} = (\textsc{mom}_2)^2$. We define the function

\begin{equation}
    f = \sum T_i (v_i - \bar{v})^2
\end{equation}

\noindent Then, the uncertainty in $f$ can be written as

\begin{align}
    \sigma_f &= \sqrt{ \sum_i^N \left[\sigma_T (v_i - v_\mathrm{mid})^2 \right]^2} \\ 
    &= \sigma_T\Delta v^2 \left(2\sum_i^{\frac{N-1}{2}} k^4\right)^{1/2}
\end{align}

\noindent We use the identity $\sum_i^n k^4 = n(n+1)(2n+1)(3n^2 + 3n -1)/30$ to obtain

\begin{align}
    \sigma_f     &= \sigma_T\Delta v^2 \frac{1}{2}\sqrt{\left(\frac{1}{15}N(N^2-1)
    \frac{1}{4}\left[3\left({N-1}\right)^2 + 3\left({N-1}\right) -4\right] \right)} \\
    &= \sigma_T\Delta v^2 \frac{1}{4} \sqrt{\left(\frac{1}{15}N(N^2-1)
    \left(3N^2-3N-4\right) \right)} \\
    &\approx \sigma_T\Delta v^2 \frac{1}{4} \sqrt{\left(\frac{3}{15}N^5 \right)} \\
    &\approx \frac{\sigma_T\Delta v^2}{4\sqrt{5}} N^2\sqrt{N} 
\end{align}

\noindent This then gives the uncertainty on the variance of

\begin{align}
    \sigma_{\textsc{var}} &= \frac{\sigma_f}{\sum T_i} \\
    &= \frac{1}{\sum T_i}\frac{\sigma_T\Delta v^2}{4\sqrt{5}} N^2\sqrt{N}  \\
    &= \frac{\sigma_I}{I}\frac{ (\Delta V_{line})^2}{4\sqrt{5}}
\end{align}

\noindent The uncertainty on the velocity dispersion is then

\begin{align}
    \sigma_{\textsc{mom}_2} &= \frac{\sigma_{\textsc{var}}}{2 {\textsc{mom}_2}} \\
    &= \frac{\sigma_I}{I}\frac{ (\Delta V_{line})^2}{4\sqrt{5}} \frac{1}{2 {\textsc{mom}_2}} \\
    &= \frac{\sigma_I}{I} \frac{(\Delta V_{line})^2}{8\sqrt{5}}\frac{1}{\textsc{mom}_2}
\end{align}

\section{Galaxy images}
\label{app:images}

\begin{figure*}
    \includegraphics[width=0.87\textwidth]{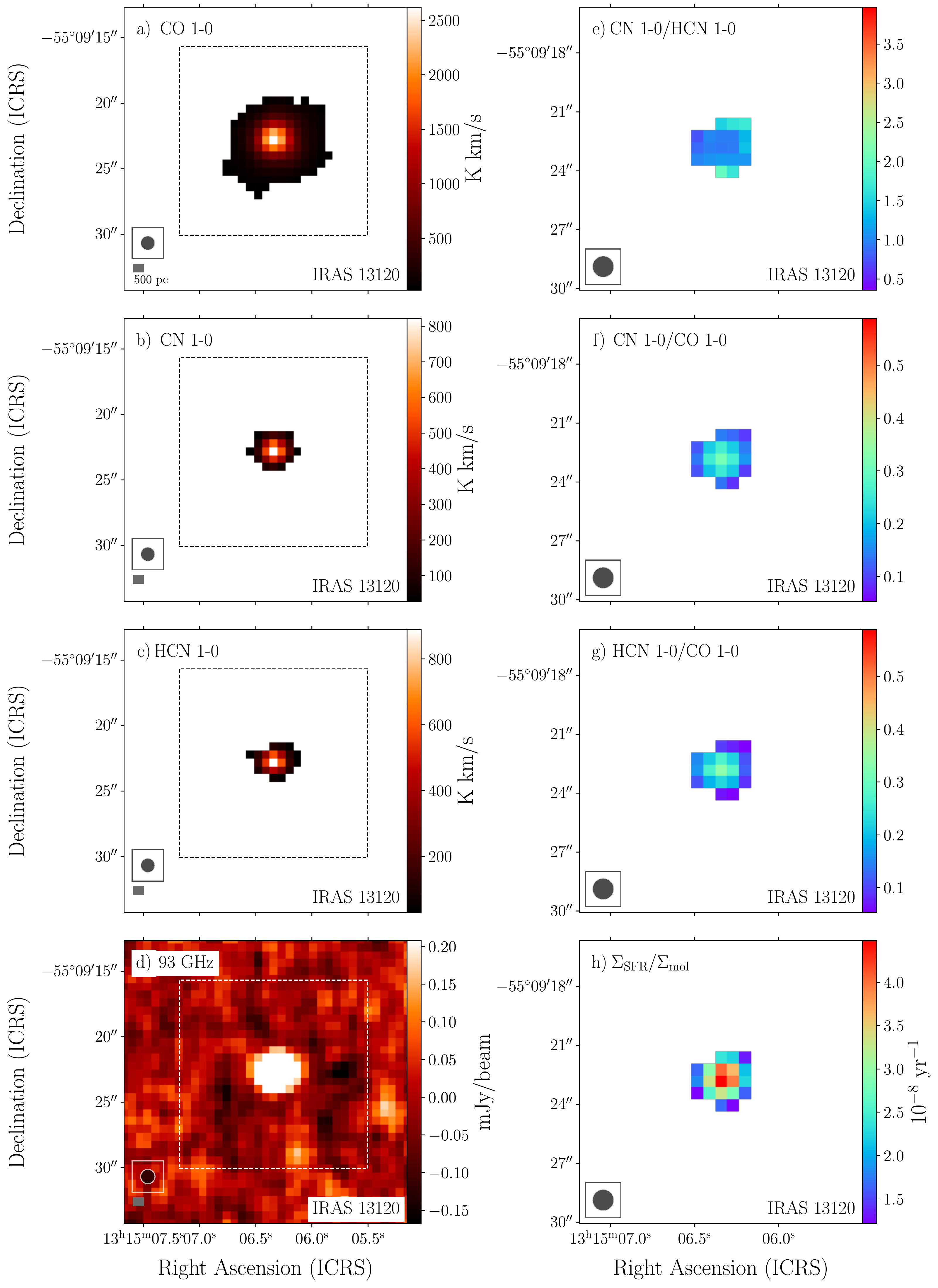}
    \caption{(Left column) Images of gas and star formation rate tracers in the ultraluminous infrared galaxy IRAS~13120: from top to bottom, CO(1-0), CN(1-0) (brighter hyperfine grouping), HCN(1-0), 93 GHz continuum. (Right column) Maps of line and continuum ratios in IRAS 13120: from top to bottom, CN/HCN ratio, CN/CO ratio, HCN/CO ratio,  $\Sigma_{SFR}/\Sigma_{\rm mol}$ ratio. Note that a smaller field of view is shown for right column (ratio images) compared to the left column (line and continuum). Only pixels with signal to noise greater than 4 in all lines and continuum are shown in the right column. The beam size is the same for all images and is indicated by the circle in the lower left corner of each image. 
    }
    \label{fig:appendixfig10}
\end{figure*}

\begin{figure*}
    \includegraphics[width=0.87\textwidth]{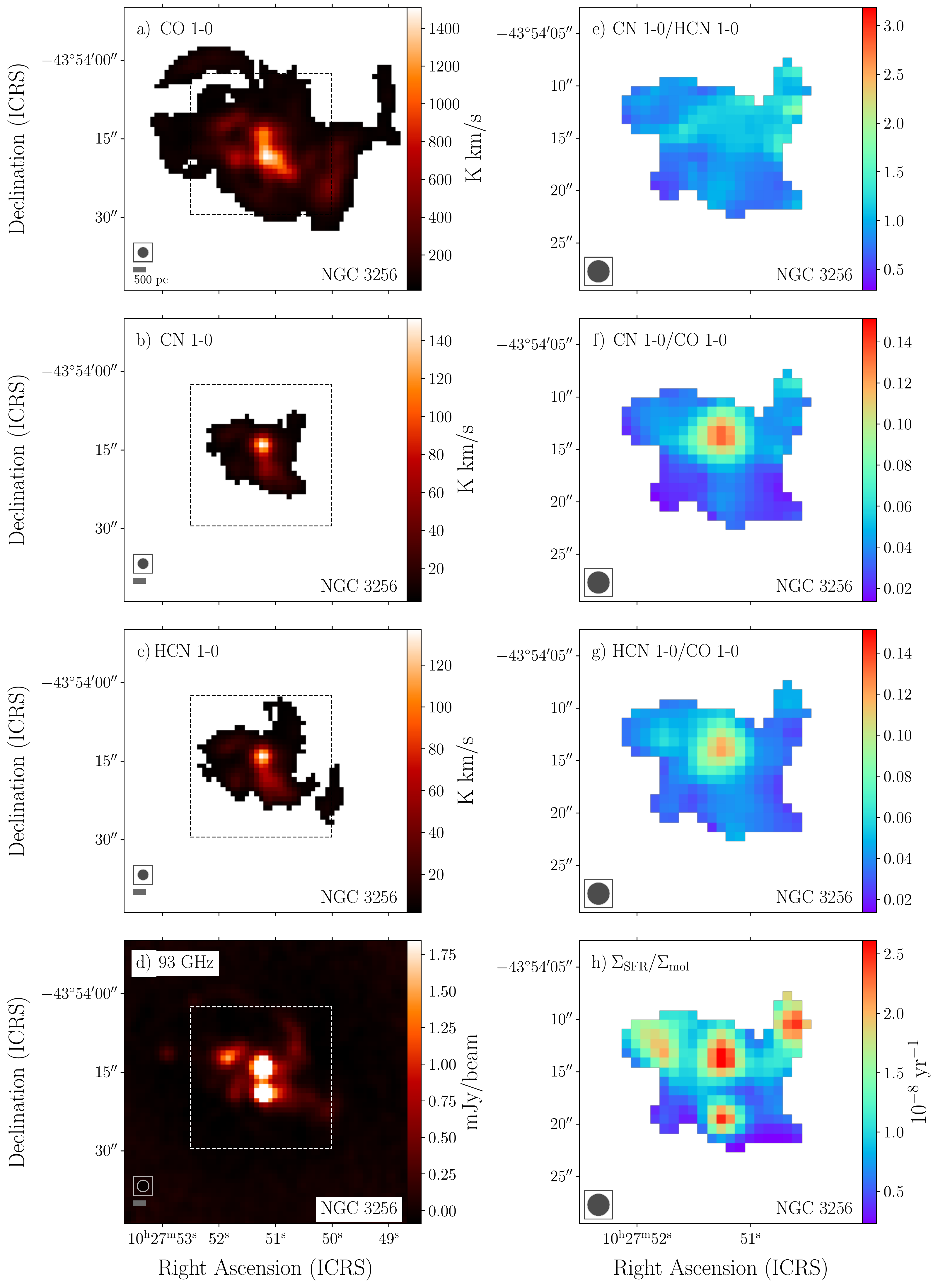}
    \caption{(Left column) Images of gas and star formation rate tracers in the luminous infrared galaxy merger NGC~3256. See Fig.~\ref{fig:appendixfig10} caption for further details.
    }
    \label{fig:appendixfig7}
\end{figure*}

\begin{figure*}
    \includegraphics[width=0.87\textwidth]{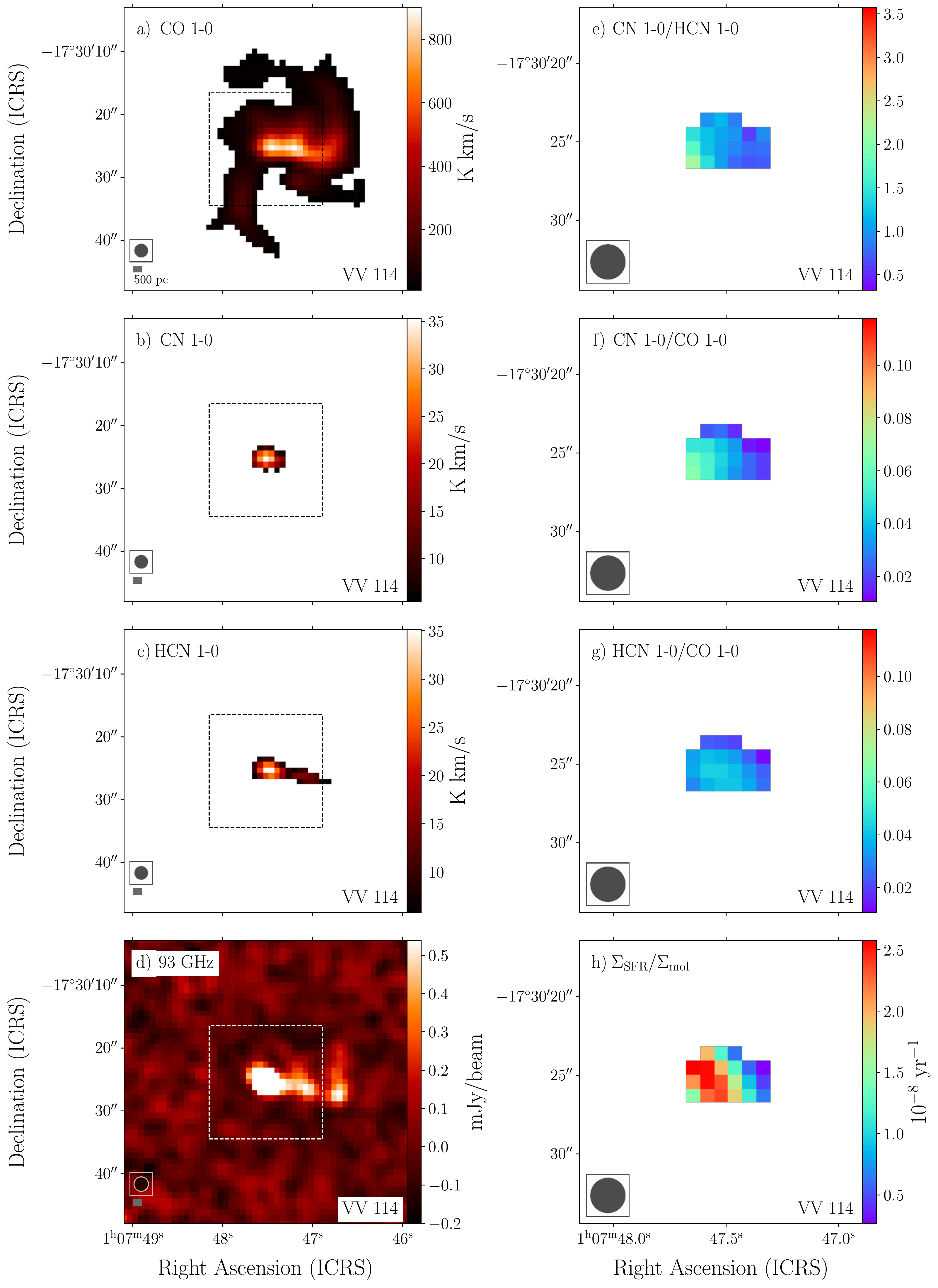}
    \caption{(Left column) Images of gas and star formation rate tracers in the luminous infrared galaxy merger VV114.  See Fig.~\ref{fig:appendixfig10} caption for further details.
    }
    \label{fig:appendixfig9}
\end{figure*}

\begin{figure*}
    \includegraphics[width=0.87\textwidth]{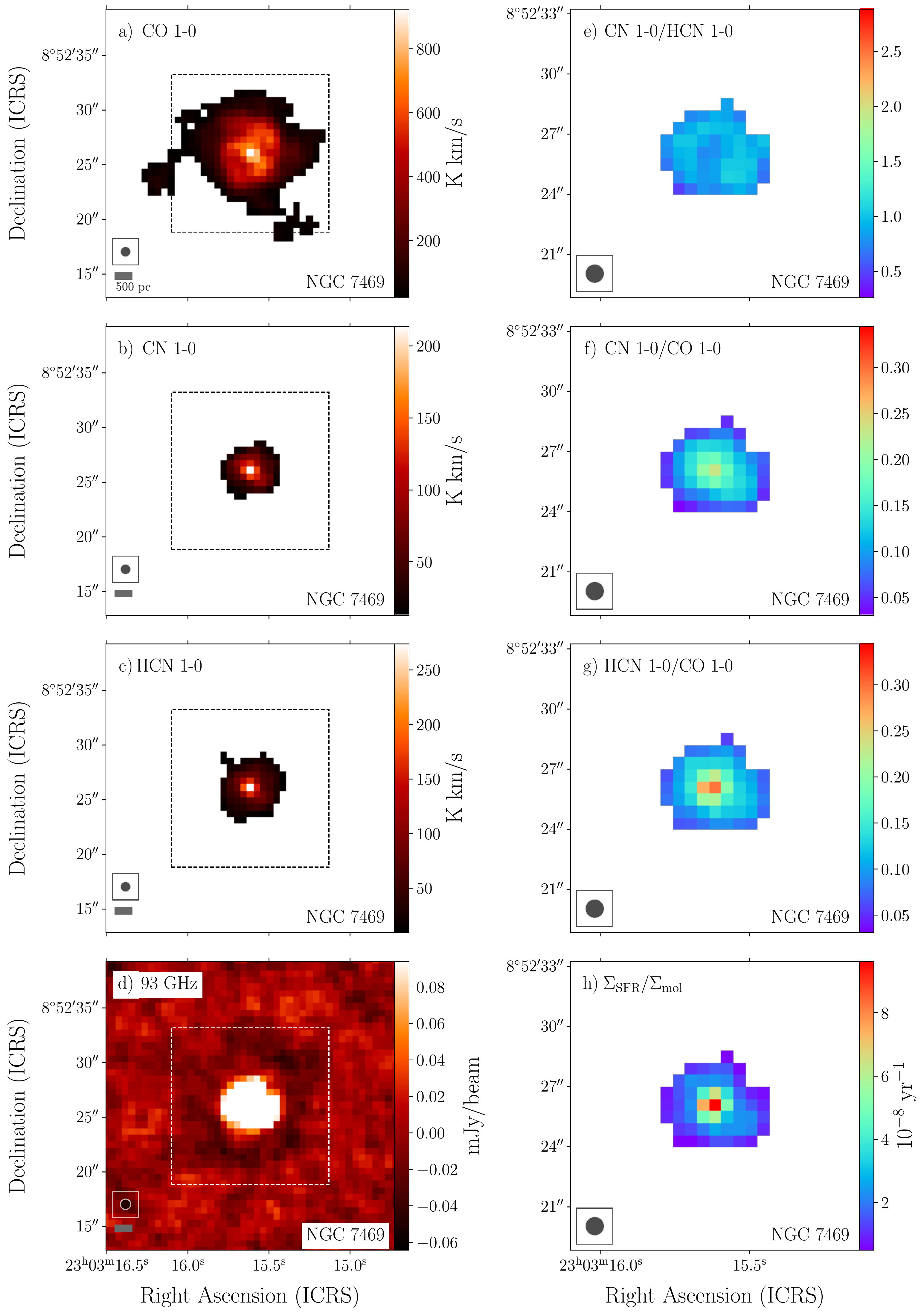}
    \caption{(Left column) Images of gas and star formation rate tracers in the luminous infrared galaxy NGC 7469.  See Fig.~\ref{fig:appendixfig10} caption for further details. The presence of a strong central AGN means that the star formation rate in the central pixels is an overestimate for this galaxy. 
    }
    \label{fig:appendixfig6}
\end{figure*}

\begin{figure*}
    \includegraphics[width=0.87\textwidth]{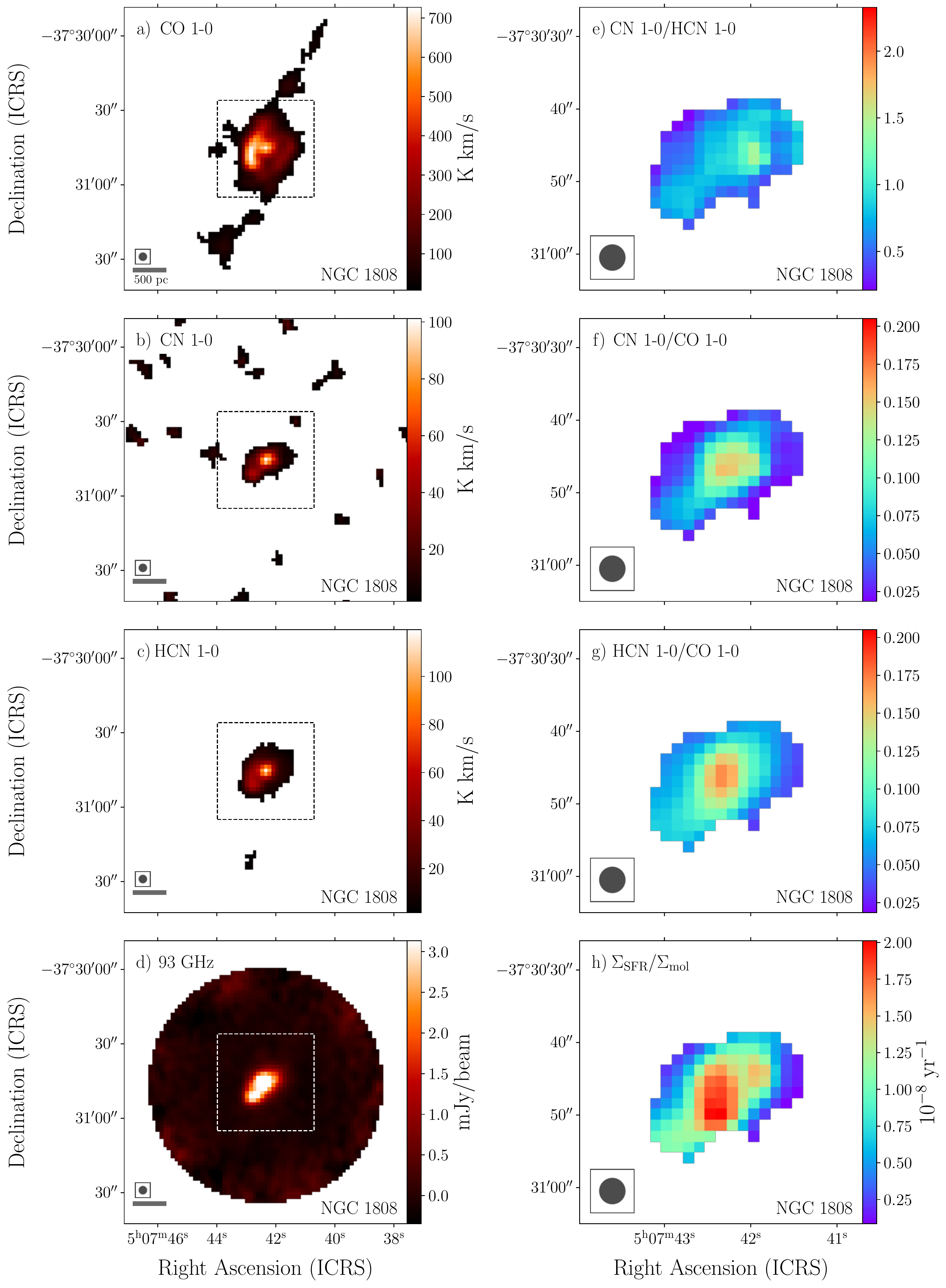}
    \caption{(Left column) Images of gas and star formation rate tracers in the galaxy NGC~1808.  See Fig.~\ref{fig:appendixfig10} caption for further details.
    }
    \label{fig:appendixfig3}
\end{figure*}

\begin{figure*}
    \includegraphics[width=0.87\textwidth]{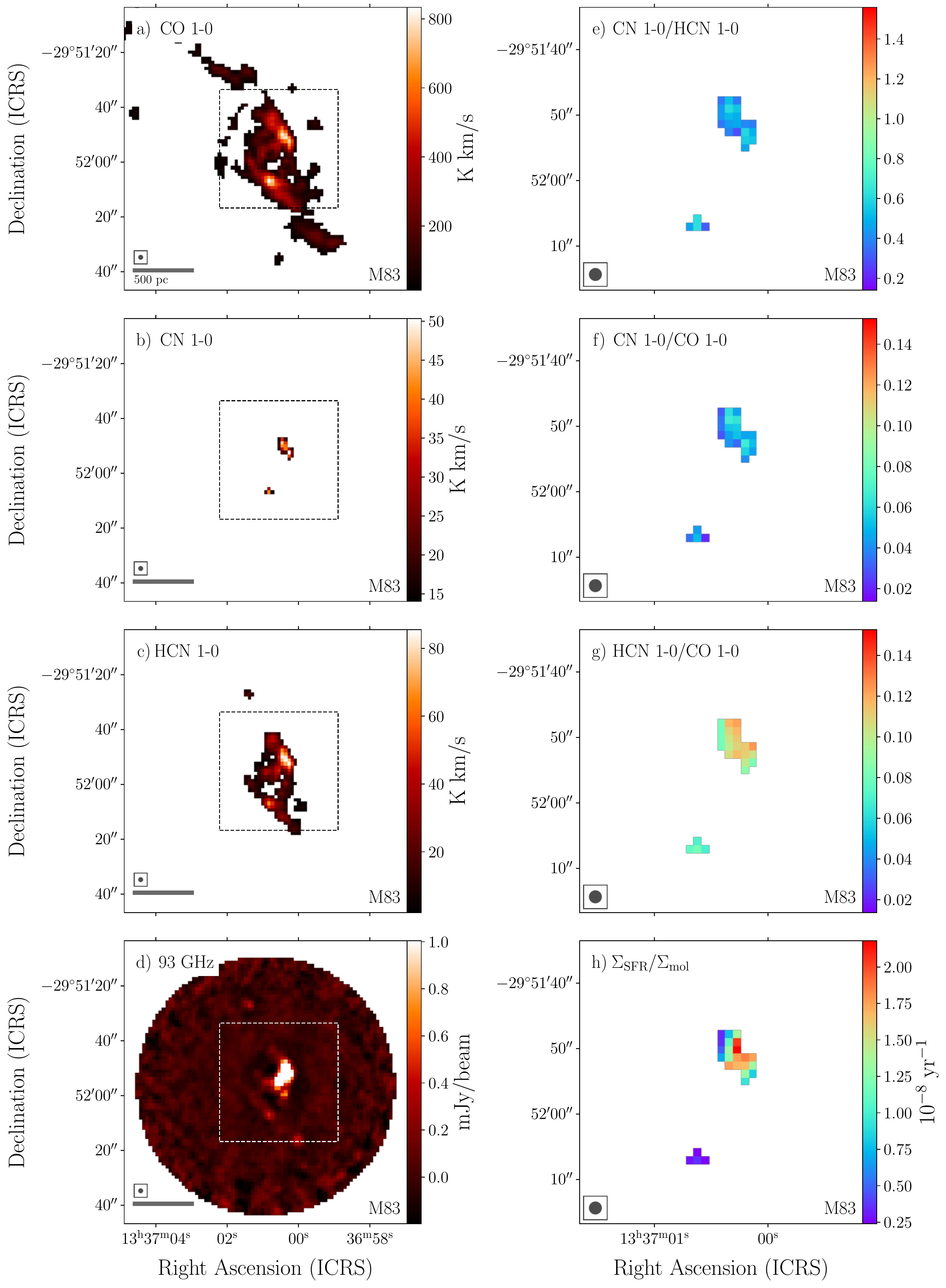}
    \caption{(Left column) Images of gas and star formation rate tracers in the galaxy M83.  See Fig.~\ref{fig:appendixfig10} caption for further details. The original CO data span a much wider area than is shown here, but CN is only detected in the central region.  
    }
    \label{fig:appendixfig8}
\end{figure*}

\begin{figure*}
    \includegraphics[width=0.87\textwidth]{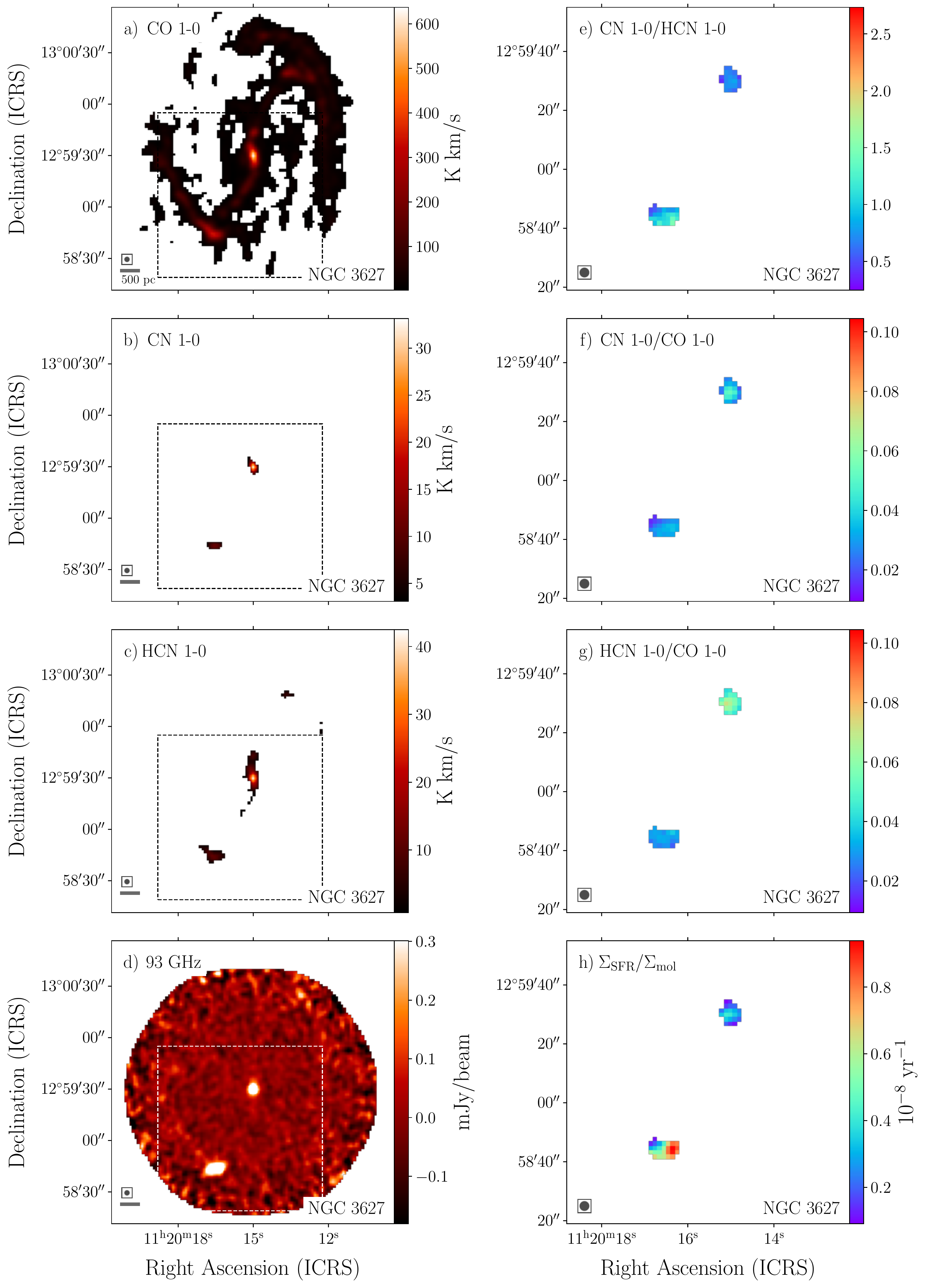}
    \caption{(Left column) Images of gas and star formation rate tracers in the galaxy NGC~3627.  See Fig.~\ref{fig:appendixfig10} caption for further details. Note that the zoomed images in the right column are not centered on the nucleus.  
    }
    \label{fig:appendixfig4}
\end{figure*}

\begin{figure*}
    \includegraphics[width=0.87\textwidth]{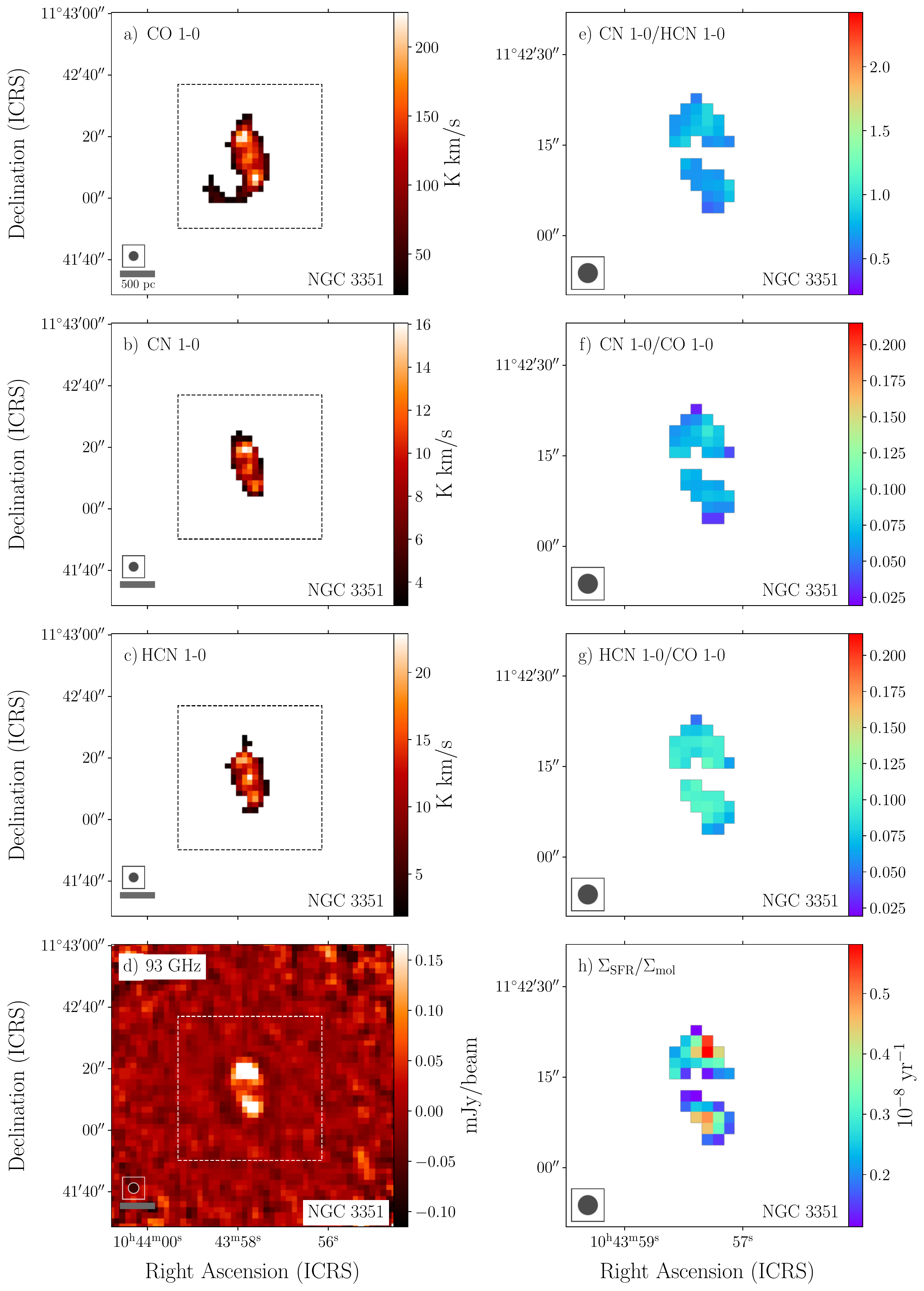}
    \caption{(Left column) Images of gas and star formation rate tracers in the galaxy NGC~3351.  See Fig.~\ref{fig:appendixfig10} caption for further details.
    }
    \label{fig:appendixfig5}
\end{figure*}

The line and continuum images and ratio maps for 8 galaxies in our sample are shown in Figures~\ref{fig:appendixfig10}-\ref{fig:appendixfig5}. The galaxies are shown in order of decreasing infrared luminosity. The corresponding images for the Circinus galaxy are shown in Figure~\ref{fig:exampleimages}.


\bsp	
\label{lastpage}
\end{document}